\documentclass[reprint, amsmath,amssymb,aps,superscriptaddress, longbibliography]{revtex4-1}
\usepackage{graphicx,braket,siunitx, hyperref}
\usepackage[dvipsnames]{xcolor}
\DeclareSIUnit\gauss{G}

\hypersetup{
    colorlinks=true,
    citecolor=Blue,
    linkcolor=Blue,
    urlcolor=Blue
}
\newcommand{\mref}[2]{%
    \hyperref[{#1}]{%
        \ref*{#1}#2%
    }%
}
\begin{document}
\title{Long rotational coherence times of molecules in a magnetic trap}

\author{L. Caldwell}
\affiliation{Centre for Cold Matter, Blackett Laboratory, Imperial College London, Prince Consort Road, London SW7 2AZ UK}

\author{H. J. Williams}
\affiliation{Centre for Cold Matter, Blackett Laboratory, Imperial College London, Prince Consort Road, London SW7 2AZ UK}

\author{N. J. Fitch}
\affiliation{Centre for Cold Matter, Blackett Laboratory, Imperial College London, Prince Consort Road, London SW7 2AZ UK}

\author{J. Aldegunde}
\affiliation{Departamento de Quimica Fisica, Universidad de Salamanca, E-37008, Salamanca, Spain}
\author{Jeremy M. Hutson}
\affiliation{Joint Quantum Centre (JQC) Durham-Newcastle, Department of Chemistry, Durham University, South Road, Durham DH1 3LE, UK}

\author{B. E. Sauer}
\affiliation{Centre for Cold Matter, Blackett Laboratory, Imperial College London, Prince Consort Road, London SW7 2AZ UK}

\author{M. R. Tarbutt}
\affiliation{Centre for Cold Matter, Blackett Laboratory, Imperial College London, Prince Consort Road, London SW7 2AZ UK}

\date{\today}

\begin{abstract}
Polar molecules in superpositions of rotational states exhibit long-range dipolar interactions, but maintaining their coherence in a trapped sample is a challenge. We present calculations that show many laser-coolable molecules have convenient rotational transitions that are exceptionally insensitive to magnetic fields. We verify this experimentally for CaF where we find a transition with sensitivity below \SI{5}{\hertz\per\gauss} and use it to demonstrate a rotational coherence time of \SI{6.4+-0.8}{\milli\second} in a magnetic trap. Simulations suggest it is feasible to extend this to more than $\SI{1}{\second}$ using a smaller cloud in a biased magnetic trap.
\end{abstract}

\maketitle
Ultracold polar molecules present exciting opportunities for quantum simulation, quantum computation, and tests of fundamental physics \cite{Carr2009}. The rotational motion is particularly important for these applications. Rotational transitions are easily driven by microwave fields, and excited rotational states have long lifetimes. Polar molecules in superpositions of rotational states have large oscillating electric dipole moments, providing long-range dipole-dipole interactions between them \cite{Barnett2006,Buchler2007,Micheli2007}. The strong electric dipole coupling of rotational states to microwave fields can also be used to interface gas-phase molecules with mesoscopic solid-state systems \cite{Andre2006,Rabl2006}. Many schemes have been proposed to use these interactions to generate entanglement, engineer inter-particle potentials and implement two-qubit quantum gates \cite{DeMille2002,Peter2012,Yao2013,Micheli2006,Karman2018}.  

Most applications require trapped molecules and coherence times that are long compared to the characteristic interaction strength. Here, we explore how that can be achieved with molecules in magnetic traps. To avoid the dephasing that arises from an inhomogeneous transition frequency, the trap potential should, ideally, be identical for the two rotational states. Experiments so far have focused exclusively on optical traps in the form of optical lattices or arrays of tweezer traps where inter-particle separations below $\SI{1}{\micro\meter}$ are possible, producing dipole-dipole couplings with energies $\sim\SI{1}{\kilo\hertz}$. Whilst second-long coherence times have been demonstrated for superpositions of hyperfine states \cite{Park2017}, rotational coherence times longer than $\sim\SI{1}{\milli\second}$ are a challenge because of the dependence of the ac Stark shift on rotational state~\cite{Kotochigova2010}. By using a dc electric field to uncouple the rotational angular momentum from the nuclear spin, and setting the polarization angle of the trapping light to equalize the Stark shifts of the chosen rotational states, \textcite{Gohle2018} extended rotational coherence times to \SI{8.7+-0.6}{\milli\second} in a sample of optically trapped NaK molecules. Long vibrational coherence times have also been demonstrated for Sr$_2$ molecules in a state-insensitive optical lattice~\cite{Kondov2019}.
 
Recently, direct laser cooling has produced ultracold molecules with both electric and magnetic moments \cite{Barry2014,Williams_2017,Anderegg2017,Collopy2018}, opening new possibilities for experiments in magnetic traps \cite{McCarron2018,Williams2018}. Two-dimensional arrays of magnetic traps have been demonstrated for atoms with spacings comparable to those achievable with optical arrays \cite{Wang2017}. Molecules might also be held in magnetic chip traps close to superconducting microwave resonators~\cite{Andre2006}, where the regime of strong coupling between molecules and microwave photons may be reached, and where molecules may be coupled via the resonator with interaction strengths of order \SI{100}{\kilo\hertz}. 
 
To obtain long rotational coherence times in a magnetic trap, it is necessary to find states with large and nearly identical magnetic moments. Here, we investigate rotational transitions in $^2\Sigma$ molecules and find a pair of states in CaF with magnetic moments of $\sim 1 \mu_\textrm{B}$ that are equal to \num{3.3+-0.1} parts per million. We demonstrate a superposition of the two states in a quadrupole magnetic trap with a coherence time of \SI{6.4+-0.8}{\milli\second}. By comparison with simulations and free-space measurements we establish the dominant dephasing mechanisms and propose ways to achieve coherence times exceeding \SI{1}{\second}.

The interaction of a $^2\Sigma$ molecule with a magnetic field $\vec{B}$ is described by the Zeeman Hamiltonian \cite{*[{}] [{. See p. 351 for Zeeman Hamiltonian and p. 635 for fine and hyperfine Hamiltonian.}] Brown2003, Bowater1973},
\begin{equation}
\begin{split}
H_{\rm Z} &= H_{\rm e} + H_{\rm n} + H_{\rm r} + H_{\rm a}\\
&= g_{S}\mu_{\rm{B}} \vec{S}\cdot \vec{B} - \sum_{i}g_{\rm{N}}^i\mu_{\rm{N}}\vec{I}^i\cdot \vec{B} - g_{\rm{r}}\mu_{\rm{B}}\vec{N}\cdot \vec{B}\\
&\quad\quad + g_l \mu_{\rm{B}}(\vec{S} \cdot \vec{B} - (\vec{S}\cdot \hat{z})(\vec{B}\cdot \hat{z})),
\end{split}
\end{equation}
where $\hat{z}$ is a unit vector in the direction of the internuclear axis, $\vec{S}$ is the electron spin operator, $\vec{N}$ is the rotational angular momentum operator, $\vec{I}^i$ is the spin operator of nucleus $i$, and the sum is over the nuclei. $H_{\rm e}$ describes the interaction of the electron magnetic moment with $\vec{B}$ and is, by far, the largest term. $H_{\rm n}$ describes the much smaller contribution from the nuclear magnetic moments, $H_{\rm r}$ the rotational Zeeman interaction, and $H_a$ the anisotropic correction to the electronic Zeeman interaction. The search for magnetically insensitive rotational transitions is hindered by the Hamiltonian for the fine and hyperfine structure, $H_{\rm fhf}$, which couples the angular momenta in a way that depends on $N$ \cite{Brown2003}. This is solved by choosing the stretched states $\ket{N}_{\mathrm{str}}=\ket{N,m_N=N}\ket{S,m_S=S}\ket{I,m_I=I}$, where $m_X$ is the projection of $X$ onto the magnetic field axis. These states are eigenstates of $H_{\rm fhf}$ and $H_{\rm Z}$. Their Zeeman shifts, $\Delta E_{N}$, are almost identical, because the large contribution from $H_{\rm e}$, and the smaller one from $H_{\rm n}$, are both independent of $N$. For the single-photon transitions $\ket{N}_{\mathrm{str}}\leftrightarrow \ket{N+1}_{\mathrm{str}}$, the residual magnetic sensitivity due to the two remaining terms is
\begin{equation}
\begin{split}
\Delta \mu(N) &= (\Delta E_{N+1} - \Delta E_N)/B, \\
&= \left(\frac{g_l}{(2N+4)^2-1}-g_{{\rm r}}\right)\mu_\textrm{B} .\label{eq:resid-sens}
\end{split}
\end{equation}
If the ratio $g_l/g_\mathrm{r}$ is close to $(2N+4)^2-1$ for some $N$, the remaining two terms nearly cancel, giving the desired magnetic insensitivity.

\begin{table*}
\caption{Results of electronic structure calculations for $g_l$ and $g_\textrm{r}$ for a variety of potentially laser-coolable molecules at their equilibrium bond lengths $R_\textrm{e}$. \label{tab:glgr}}
\centering
\begin{ruledtabular}
\begin{tabular}{@{}l *{12}{r}@{}}
Molecule          & BeF & MgF & CaF & SrF & BeH & MgH & CaH & SrH & BeOH
& MgOH & CaOH & SrOH \\
\hline
$R_\textrm{e}$ (M--X) (\AA) & 1.36  & 1.75  & 1.95 &2.07 & 1.34 & 1.73 & 2.00 & 2.15 & 1.37&1.76 &2.03 & 2.16 \\
$R_\textrm{e}$ (O--H) (\AA) & --- & --- &  ---    &  ---   & --- & --- & --- &--- & 0.94& 0.94& 0.95& 0.97 \\
$10^3 g_l$      & $-0.820$ & $-1.74$ & $-1.86$ & $-4.97$ & $-0.111$ & $-2.18$ & $-4.20$ & $-15.1$ & $-0.394$& $-1.26$ & $-1.73$ & $-5.24$ \\
$10^5 g_\textrm{r}$ & $-$7.36 & $-3.73$ & $-5.13$ & $-4.77$ & $-144$ & $-88.5$ & $-106$ & $-111$ & $-2.18$& $-1.48$ & $-3.31$ & $-3.17$ \\
$g_{l}/g_\textrm{r}$ & 11.1 & 46.6 & 36.3 & 104 & 0.077 & 2.46 & 3.95 & 13.6 & 18.0 & 85.0 & 52.3 & 165\\
\end{tabular}
\end{ruledtabular}
\end{table*}

We have carried out calculations of this ratio for a variety of alkaline-earth fluorides, hydrides and hydroxides of interest for laser cooling. To evaluate $g_l=\Delta g_{\bot}=g_{\|}-g_{\bot}$, we calculate the parallel and perpendicular components of the molecular $g$ tensor by density-functional theory (DFT) as implemented in ORCA \cite{Neese:2012}, using the B3LYP functional \cite{Stephens:1994} and x2c-TZVPP all-electron basis sets \cite{Pollak:2017}. Relativistic corrections are included by the zero-order regular approximation (ZORA) \cite{vanLenthe:1993, vanLenthe:1994,vanLenthe:1999}, but are small for the molecules considered here. We evaluate $g_\textrm{r}$ from Hartree-Fock calculations, as implemented in DALTON \cite{Dalton:current}, using the same basis sets. Further details are given in the Supplemental Materials \cite{Supp}.\nocite{aldegunde:2Sigma:pra2019,curl:mp1965, Anderson:JCP1994:MgF,Childs:JMS1981:SrF,Knight:JCP1972:BeH,Bernath:TAJ1985:MgH,Berg1974:CaH,:SrH,Brom:JCP1976:BeOH,Brom:JCP:1973:MgOH,Ziurys:TAJ1992:CaOH,Anderson:CPL1992:SrOH,Weltner:book,Mohn:AQC2005,Lee:T1}

\begin{figure}
    \centering
    \includegraphics[width=\linewidth]{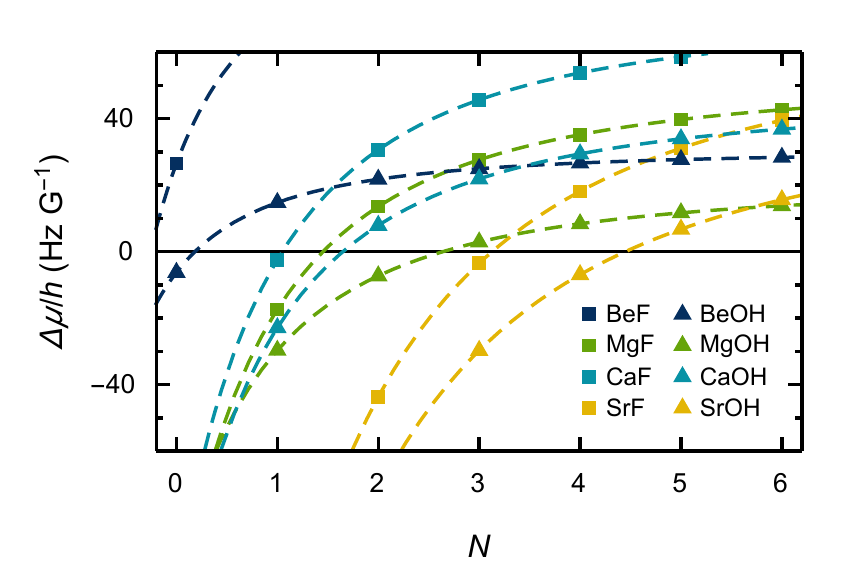}
    \caption{Calculated magnetic sensitivities of transitions $\ket{N}_{\mathrm{str}}\leftrightarrow \ket{N+1}_{\mathrm{str}}$ for a variety of alkaline-earth fluorides and hydroxides of interest for laser cooling.}
    \label{fig:theoryPred}
\end{figure}

The results obtained when the bond length for each molecule is fixed at its equilibrium value $R_\textrm{e}$ are summarized in Table \ref{tab:glgr}. Figure~\ref{fig:theoryPred} shows the resulting $\Delta \mu$ for transitions $\ket{N}_{\mathrm{str}}\leftrightarrow \ket{N+1}_{\mathrm{str}}$ up to $N=6$. For $N=0$, the sensitivities are dominated by $g_l$, and for large $N$ they approach the value set by $g_{\rm r}$. The hydrides generally have much larger values of both $g_\textrm{r}$ and $g_l$ and have no transitions with residual sensitivities on the scale shown. The remaining molecules each exhibit at least one transition with $|\Delta \mu /h|<\SI{30}{\hertz\per\gauss}$ and six of them have a transition with $|\Delta \mu /h|<\SI{10}{\hertz\per\gauss}$. These results suggest that convenient rotational transitions with extremely small magnetic sensitivities are a common feature of alkaline-earth fluorides and hydroxides.

\begin{figure}
    \centering
    \includegraphics[width=\linewidth]{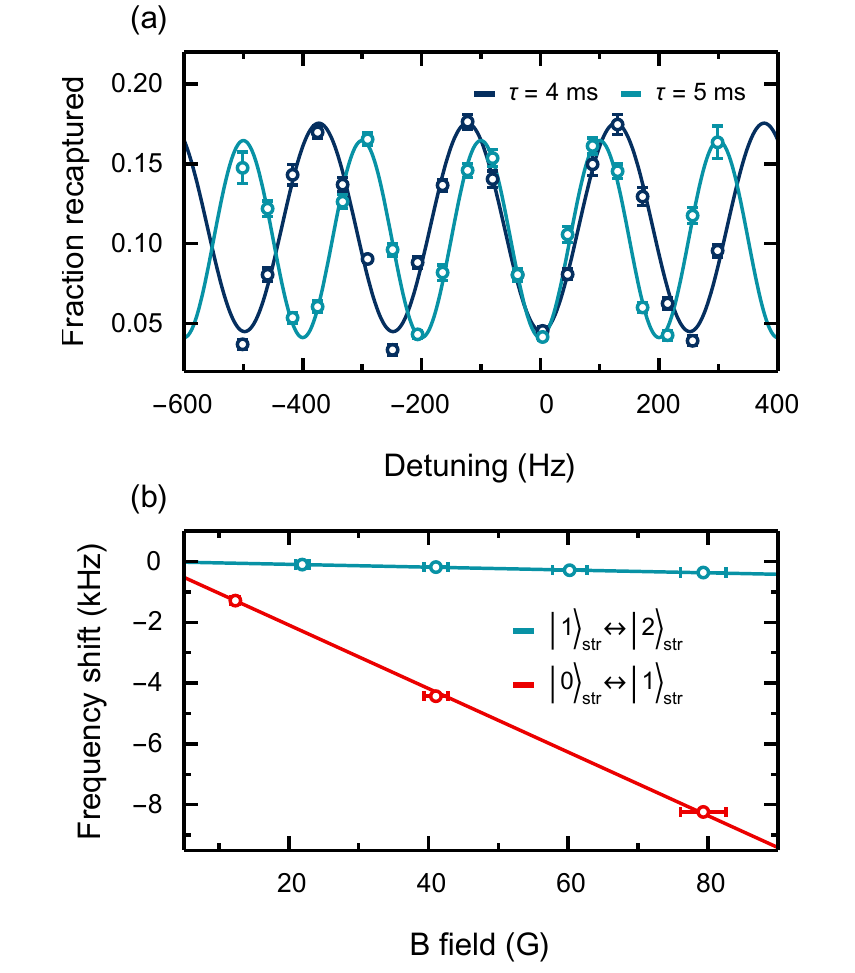}
    \caption{(a) Ramsey spectroscopy of the transition $\ket{1}_{\mathrm{str}}\leftrightarrow\ket{2}_{\mathrm{str}}$ in free space with $\tau=4$ and \SI{5}{\milli\second}. Points and error bars: mean and standard error of 20 repeated experiments; lines: sinusoidal fits to the data. (b) Magnetic sensitivity of the transitions $\ket{0}_{\mathrm{str}}\leftrightarrow\ket{1}_{\mathrm{str}}$ (red) and $\ket{1}_{\mathrm{str}}\leftrightarrow\ket{2}_{\mathrm{str}}$ (light blue). Lines: linear fits to the data. Vertical error bars are smaller than the points.}
    \label{fig:magSens}
\end{figure}

We have used Ramsey spectroscopy in a uniform magnetic field $B$ to determine the magnetic sensitivities of such transitions experimentally in CaF. We measure $\Delta \mu$ for two transitions, $\ket{0}_{\mathrm{str}}\leftrightarrow \ket{1}_{\mathrm{str}}$ and $\ket{1}_{\mathrm{str}}\leftrightarrow \ket{2}_{\mathrm{str}}$, whose frequencies are $f_{0\leftrightarrow1}=\SI{20553.4}{\mega\hertz}$ and  $f_{1\leftrightarrow2}=\SI{41088.9}{\mega\hertz}$. We capture $\sim 10^4$ molecules in a magneto-optical trap \cite{Williams_2017}, cool them to about \SI{50}{\micro\kelvin}~\cite{Truppe2017b}, then prepare them in a single internal state, either $\ket{0}_{\mathrm{str}}$ or $\ket{1}_{\mathrm{str}}$~\cite{Williams2018}. The Ramsey sequence \cite{Blackmore2018} uses a pair of nearly resonant $\pi/2$ pulses, \SI{20}{\micro\second} long, separated by a free-evolution time $\tau$. Following the second pulse, we measure the number of molecules in $N=1$ by recapturing them into the MOT and imaging their fluorescence.  Dividing by the number initially loaded in the MOT gives a signal that is insensitive to shot-to-shot fluctuations in the number of molecules. The magnetic field at the molecules is calibrated to better than 1\% by single-pulse microwave spectroscopy of the transition $\ket{N=0,F=1,m_F=-1}\leftrightarrow\ket{N=1,F=0}$. The Zeeman shift of this transition can be accurately calculated because it is dominated by $H_{\rm e}$, and because $g_{S}$ and the hyperfine parameters are known to high precision~\cite{Childs:JMS1981:CaF}. 

Figure~\mref{fig:magSens}{(a)} shows Ramsey fringes for the transition $\ket{1}_{\mathrm{str}}\leftrightarrow\ket{2}_{\mathrm{str}}$, obtained by scanning the microwave frequency. The dark blue data are for $\tau=4$~ms and the light blue for $\tau=5$~ms. Repeating the measurement with different $\tau$ identifies the fringe corresponding to the centre frequency. Fitting to the fringes determines the transition frequency with statistical uncertainty below \SI{1}{\hertz}.

Figure~\mref{fig:magSens}{(b)} shows the change in transition frequency for both transitions as a function of $B$. The solid lines show linear fits to the data that give residual magnetic sensitivities $\Delta\mu/h=\SI{-104+-4}{\hertz\per\gauss}$ for the transition $\ket{0}_{\mathrm{str}}\leftrightarrow\ket{1}_{\mathrm{str}}$ and \SI{-4.7+-0.2}{\hertz\per\gauss} for $\ket{1}_{\mathrm{str}}\leftrightarrow\ket{2}_{\mathrm{str}}$. The uncertainties are dominated by drifts in $B$ between measurements of the insensitive transition and the transition used for calibration, represented by the horizontal error bars. The error in magnetic field is correlated between different points, because the drift is slow compared to the measurement time; this has been taken into account in the calculation of the uncertainties in $\Delta\mu$. From these measurements we determine $g_l=\num{-1.87+-0.08e-3}$ and $g_{\rm r}=\num{-5.0+-0.2e-5}$, in excellent agreement with the calculated values. At $N=1$ the two terms in Eq.~\ref{eq:resid-sens} cancel to within 3\%, resulting in cancellation of the magnetic moments of the two states to \num{3.3+-0.1} parts per million.

\begin{figure}
    \centering
    \includegraphics[width=\linewidth]{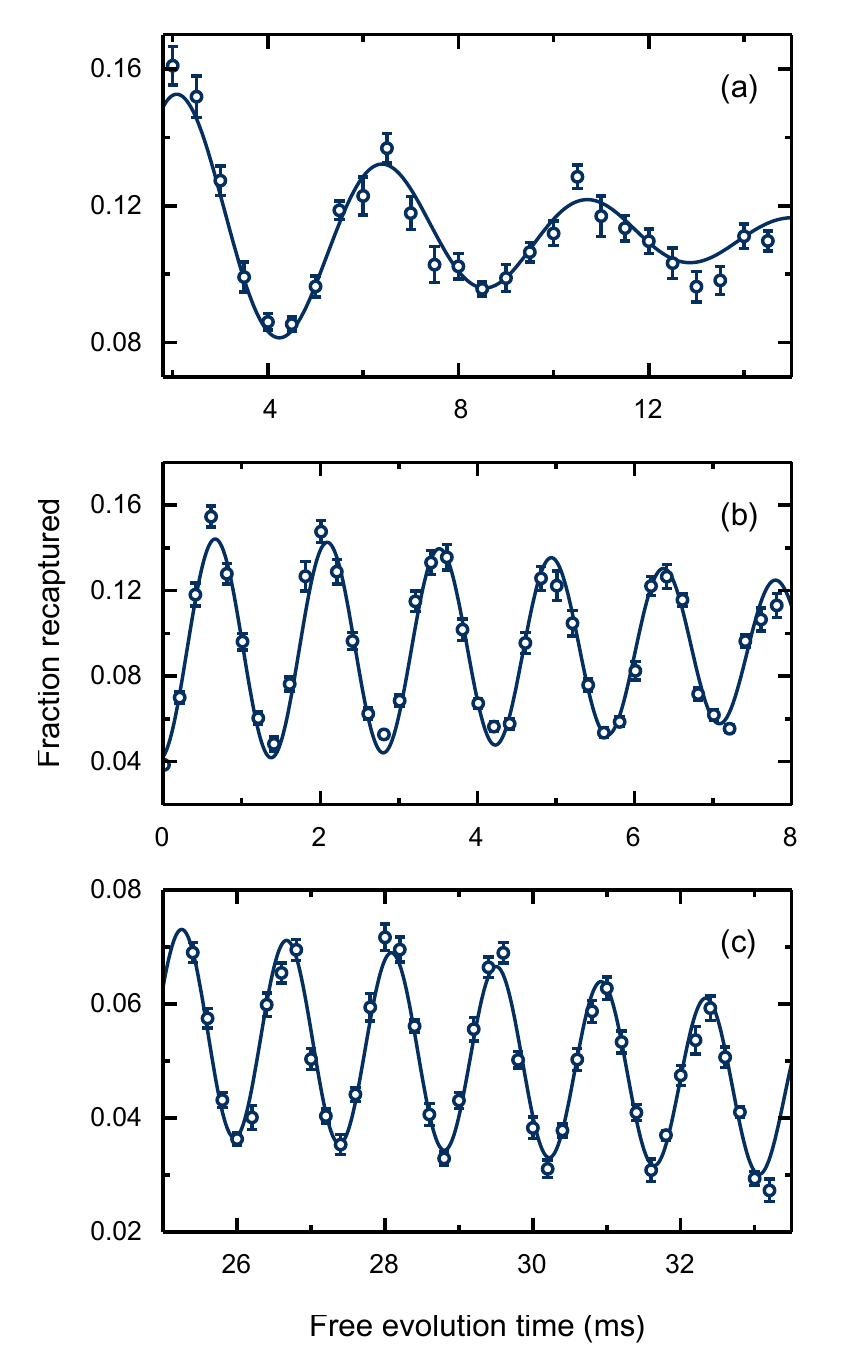}
    \caption{(a) Ramsey fringes for molecules in the magnetic trap. Line: fit to $A + B e^{- \alpha t} \cos(\Delta t + \phi)$ giving a coherence time $1/\alpha=\SI{6.4+-0.8}{\milli\second}$. (b) Ramsey fringes for molecules in free space. Line: fit to $A + B e^{-\alpha^2 t^2} \cos (\Delta t + \phi)$, with $1/\alpha = (\lambda/2\pi)\sqrt{m/2 k_{\rm B} T}=\SI{11.4+-0.6}{\milli\second}$, corresponding to $T=\SI{37+-4}{\micro\kelvin}$. (c) As (b) but with a spin echo applied: an extra $\pi$ pulse is applied, \SI{14.7}{\milli\second} after the first $\pi/2$ pulse. The $x$ axis shows total free-evolution time between first and second $\pi/2$ pulses. Line: fit to $(1-C t^2)(A + B e^{-\alpha t} \cos(\Delta t + \phi))$. The loss of fringe contrast is consistent with $\alpha=0$. Points and error bars throughout show mean and standard error of 20 repeated experiments.}
    \label{fig:coherence}
\end{figure}

The magnetic insensitivity of the transition $\ket{1}_\mathrm{str}\leftrightarrow\ket{2}_\mathrm{str}$ suggests that long coherence times should be possible in a magnetic trap. To investigate this we measure the decay of the beat note between the oscillations of the rotational superposition and those of a slightly detuned microwave field. The first $\pi/2$ pulse is applied and then the quadrupole trap is immediately turned on with an axial field gradient of \SI{45}{\gauss\per\centi\meter}. The lifetime of the molecules in the magnetic trap is about 4.5~s~\cite{Williams2018}, much longer than the timescale of any of the experiments reported here. The second $\pi/2$ pulse is applied in the trap and the molecules are then recaptured and imaged in a MOT, as before. Figure~\mref{fig:coherence}{(a)} shows the Ramsey fringes traced out by scanning the free evolution time $\tau$. The solid line is a fit to a sinusoid with exponentially decaying amplitude, which gives a coherence time of \SI{6.4+-0.8}{\milli\second}.

To elucidate the sources of decoherence in the trap, we use a Monte-Carlo simulation of the Ramsey experiment in the magnetic trap. The result of this simulation fits well to an exponentially decaying sinusoid with $1/e$ time of \SI{6.1}{\milli\second}, in agreement with our experimental data. We identify three decoherence mechanisms: (i) The residual magnetic sensitivity of the transition causes a spread in the phase accumulated by different molecules. (ii) The movement of a molecule between the two $\pi/2$ pulses causes a change of phase which is different for each molecule; this is a form of Doppler broadening. (iii) The molecules adiabatically follow the local magnetic field vector as they move along different trajectories. This last mechanism has two effects. The first is to vary the amplitude of the microwave field in the relevant polarisation for each molecule at each Ramsey pulse, resulting in a loss of contrast, but no longer-term dephasing. The second is to impart a geometric phase to each molecule, which depends on its trajectory between the two pulses, and is proportional to the difference in $m_F$ between the two states \cite{[{See for example section 5.2 of }]Tarbutt2009}. To quantify the contributions of each mechanism, we artificially remove the effect of the others from the simulation and find the time taken for the amplitude of the Ramsey fringes to decay by a factor of $e$. We find the decoherence rates for each mechanism to be: (i) \SI{60}{\per\second}, (ii) \SI{50}{\per\second} and (iii) \SI{100}{\per\second}.

The decoherence rate of mechanism (i) scales linearly with the average field experienced by the molecules in the trap, currently set almost entirely by the initial Gaussian width of \SI{1.5}{\milli\meter}. In the limit of a small initial cloud, the sample explores a range of fields determined only by its temperature, $T$. The resulting spread of transition frequencies, $(k_{\rm B} T/h)(\Delta\mu/\mu_{\rm_B})$, gives a decoherence rate of \SI{0.14}{\per\micro\kelvin\per\second}. This implies an achievable coherence time of \SI{1.4}{\second} at \SI{5}{\micro\kelvin}, a temperature that has already been demonstrated for CaF \cite{Caldwell2019, Cheuk2018}. The effect of mechanism (ii) is also related to the cloud size. When the maximum distance that a molecule moves is less than the wavelength of the microwave field, the phase change due to motion is limited to less than $2\pi$ and the superposition never fully decoheres. The result is a fixed loss of contrast at late times. In our experiment, this maximum phase change is set by the initial cloud size; reducing the cloud size would decrease the size of the effect. Mechanism (iii), currently the largest contribution to the decoherence rate, can be greatly reduced by using a magnetic trap geometry with a large bias field \cite{Bergeman1987} to ensure that there is little variation of the magnetic field direction in the region where the molecules are confined.

To identify any decoherence mechanisms unrelated to the trap, we repeated the decoherence measurement in free space, as shown in Fig.~\mref{fig:coherence}{(b)}. The data fit well to a model where the only source of decoherence is the expansion of the cloud along the $k$ vector of the microwaves i.e. mechanism (ii). In free space, the velocity of each molecule along the $k$ vector is constant so the phase accumulated changes linearly with time. Consequently, this decoherence can be reversed by employing a spin-echo sequence in which a $\pi$ pulse is applied mid-way between the two $\pi/2$ pulses. Figure~\mref{fig:coherence}{(c)} shows Ramsey fringes in free space with a $\pi$ pulse applied 14.7~ms after the first $\pi/2$ pulse. Molecules at different positions experience different microwave powers, so a $\pi$ or $\pi/2$ pulse cannot be perfect for all molecules. The extra $\pi$ pulse for the spin-echo technique thus causes an overall reduction in contrast. The signal also falls as the molecules expand and drop out of the detection region. However, we see no decoherence at all on this timescale; our data restrict the magnitude of any effect to less than 20~s$^{-1}$ at the 95\% confidence level.

The exceptionally small magnetic sensitivity of the transition $\ket{1}_\mathrm{str}\leftrightarrow\ket{2}_\mathrm{str}$ raises the question of whether the sensitivity might be even smaller in an excited vibrational state, $v>0$. To consider this, we expand the calculated $g_l$ and $g_\textrm{r}$ for CaF as a Taylor series in the bond length $R$ around the equilibrium value $R_\textrm{e}$. For a harmonic oscillator, only even-order terms in this expansion contribute to expectation values. For a slightly anharmonic oscillator such as most chemically bound molecules, both the linear and quadratic terms contribute significantly to the dependence on $v$. We calculate $g_l$ and $g_\textrm{r}$ at two additional bond lengths, close to the inner and outer turning points for the $v=1$ state of CaF, to evaluate the coefficients linear and quadratic in $(R-R_\textrm{e})$. Combining these with the expectation values of $(R-R_\textrm{e})$ and $(R-R_\textrm{e})^2$ for CaF yields
\begin{align}
&g_l(v)=-1.86\times10^{-3}\left[1+6.20\times10^{-3}\left(v+\textstyle{\frac{1}{2}}\right) + \ldots\right];\nonumber\\
&g_\textrm{r}(v)=-5.13\times10^{-5}\left[1-5.77\times10^{-3}\left(v+\textstyle{\frac{1}{2}}\right)+\ldots\right];\nonumber\\
&g_l(v)/g_\textrm{r}(v)=36.16\left[1+0.0120\left(v+\textstyle{\frac{1}{2}}\right)+\ldots\right].
\end{align}
This indicates that the dependence on $v$ is small, and that the magnetic sensitivity is lowest for $v=0$. Nevertheless, the magnitude of the effect suggests that vibrational state dependence may be useful for fine-tuning magnetic sensitivity in other molecules.

In conclusion, we have shown that many laser-coolable molecules have pairs of states belonging to neighbouring rotational manifolds whose magnetic moments are large and identical to a few parts per million. In CaF, the transition between the stretched states of $N=1$ and $N=2$ has a magnetic sensitivity of \SI{-4.7+-0.2}{\hertz\per\gauss}. We have demonstrated a coherence time of \SI{6.4+-0.8}{\milli\second} for a magnetically trapped sample in a superposition of these states. Characterisation of the principal sources of decoherence suggests that modifications to the trapping geometry and reduction of the cloud size will yield considerable improvements, with the prospect of coherence times greater than \SI{1}{\second} for small clouds at \SI{5}{\micro\kelvin} or below. These results open up potential new platforms for quantum simulation and computation with polar molecules in magnetic micro-traps \cite{Wang2017} and provide a route to the regime of strong coupling between solid-state systems and molecules held in magnetic chip traps \cite{Andre2006}. These magnetically insensitive transitions may also be valuable in precision measurement \cite{[{See for example }]Kajita2008} where sensitivity to magnetic fields is an important source of systematic error.

Underlying data may be accessed from Zenodo \cite{zenodo} and used under the Creative Commons CCZero license. We thank J. Dyne and V. Gerulis for expert technical assistance. This work was supported by EPSRC under grants EP/M027716/1 and EP/P01058X/1, and by the Royal Society.

L. Caldwell and H.J. Williams contributed equally to this work.
\bibliography{library,Sup_Mat}

\begin{thebibliography}{56}%
\makeatletter
\providecommand \@ifxundefined [1]{%
 \@ifx{#1\undefined}
}%
\providecommand \@ifnum [1]{%
 \ifnum #1\expandafter \@firstoftwo
 \else \expandafter \@secondoftwo
 \fi
}%
\providecommand \@ifx [1]{%
 \ifx #1\expandafter \@firstoftwo
 \else \expandafter \@secondoftwo
 \fi
}%
\providecommand \natexlab [1]{#1}%
\providecommand \enquote  [1]{``#1''}%
\providecommand \bibnamefont  [1]{#1}%
\providecommand \bibfnamefont [1]{#1}%
\providecommand \citenamefont [1]{#1}%
\providecommand \href@noop [0]{\@secondoftwo}%
\providecommand \href [0]{\begingroup \@sanitize@url \@href}%
\providecommand \@href[1]{\@@startlink{#1}\@@href}%
\providecommand \@@href[1]{\endgroup#1\@@endlink}%
\providecommand \@sanitize@url [0]{\catcode `\\12\catcode `\$12\catcode
  `\&12\catcode `\#12\catcode `\^12\catcode `\_12\catcode `\%12\relax}%
\providecommand \@@startlink[1]{}%
\providecommand \@@endlink[0]{}%
\providecommand \url  [0]{\begingroup\@sanitize@url \@url }%
\providecommand \@url [1]{\endgroup\@href {#1}{\urlprefix }}%
\providecommand \urlprefix  [0]{URL }%
\providecommand \Eprint [0]{\href }%
\providecommand \doibase [0]{http://dx.doi.org/}%
\providecommand \selectlanguage [0]{\@gobble}%
\providecommand \bibinfo  [0]{\@secondoftwo}%
\providecommand \bibfield  [0]{\@secondoftwo}%
\providecommand \translation [1]{[#1]}%
\providecommand \BibitemOpen [0]{}%
\providecommand \bibitemStop [0]{}%
\providecommand \bibitemNoStop [0]{.\EOS\space}%
\providecommand \EOS [0]{\spacefactor3000\relax}%
\providecommand \BibitemShut  [1]{\csname bibitem#1\endcsname}%
\let\auto@bib@innerbib\@empty
\bibitem [{\citenamefont {Carr}\ \emph {et~al.}(2009)\citenamefont {Carr},
  \citenamefont {DeMille}, \citenamefont {Krems},\ and\ \citenamefont
  {Ye}}]{Carr2009}%
  \BibitemOpen
  \bibfield  {author} {\bibinfo {author} {\bibfnamefont {L.~D.}\ \bibnamefont
  {Carr}}, \bibinfo {author} {\bibfnamefont {D.}~\bibnamefont {DeMille}},
  \bibinfo {author} {\bibfnamefont {R.~V.}\ \bibnamefont {Krems}}, \ and\
  \bibinfo {author} {\bibfnamefont {J.}~\bibnamefont {Ye}},\ }\bibfield
  {title} {\enquote {\bibinfo {title} {Cold and ultracold molecules: science,
  technology and applications},}\ }\href {\doibase
  10.1088/1367-2630/11/5/055049} {\bibfield  {journal} {\bibinfo  {journal}
  {New J. Phys.}\ }\textbf {\bibinfo {volume} {11}},\ \bibinfo {pages} {055049}
  (\bibinfo {year} {2009})}\BibitemShut {NoStop}%
\bibitem [{\citenamefont {Barnett}\ \emph {et~al.}(2006)\citenamefont
  {Barnett}, \citenamefont {Petrov}, \citenamefont {Lukin},\ and\ \citenamefont
  {Demler}}]{Barnett2006}%
  \BibitemOpen
  \bibfield  {author} {\bibinfo {author} {\bibfnamefont {R.}~\bibnamefont
  {Barnett}}, \bibinfo {author} {\bibfnamefont {D.}~\bibnamefont {Petrov}},
  \bibinfo {author} {\bibfnamefont {M.}~\bibnamefont {Lukin}}, \ and\ \bibinfo
  {author} {\bibfnamefont {E.}~\bibnamefont {Demler}},\ }\bibfield  {title}
  {\enquote {\bibinfo {title} {Quantum magnetism with multicomponent dipolar
  molecules in an optical lattice},}\ }\href {\doibase
  10.1103/PhysRevLett.96.190401} {\bibfield  {journal} {\bibinfo  {journal}
  {Phys. Rev. Lett.}\ }\textbf {\bibinfo {volume} {96}},\ \bibinfo {pages}
  {190401} (\bibinfo {year} {2006})}\BibitemShut {NoStop}%
\bibitem [{\citenamefont {B\"uchler}\ \emph {et~al.}(2007)\citenamefont
  {B\"uchler}, \citenamefont {Demler}, \citenamefont {Lukin}, \citenamefont
  {Micheli}, \citenamefont {Prokof'ev}, \citenamefont {Pupillo},\ and\
  \citenamefont {Zoller}}]{Buchler2007}%
  \BibitemOpen
  \bibfield  {author} {\bibinfo {author} {\bibfnamefont {H.~P.}\ \bibnamefont
  {B\"uchler}}, \bibinfo {author} {\bibfnamefont {E.}~\bibnamefont {Demler}},
  \bibinfo {author} {\bibfnamefont {M.}~\bibnamefont {Lukin}}, \bibinfo
  {author} {\bibfnamefont {A.}~\bibnamefont {Micheli}}, \bibinfo {author}
  {\bibfnamefont {N.}~\bibnamefont {Prokof'ev}}, \bibinfo {author}
  {\bibfnamefont {G.}~\bibnamefont {Pupillo}}, \ and\ \bibinfo {author}
  {\bibfnamefont {P.}~\bibnamefont {Zoller}},\ }\bibfield  {title} {\enquote
  {\bibinfo {title} {{Strongly Correlated 2D Quantum Phases with Cold Polar
  Molecules: Controlling the Shape of the Interaction Potential}},}\ }\href
  {\doibase 10.1103/PhysRevLett.98.060404} {\bibfield  {journal} {\bibinfo
  {journal} {Phys. Rev. Lett.}\ }\textbf {\bibinfo {volume} {98}},\ \bibinfo
  {pages} {060404} (\bibinfo {year} {2007})}\BibitemShut {NoStop}%
\bibitem [{\citenamefont {Micheli}\ \emph {et~al.}(2007)\citenamefont
  {Micheli}, \citenamefont {Pupillo}, \citenamefont {B\"uchler},\ and\
  \citenamefont {Zoller}}]{Micheli2007}%
  \BibitemOpen
  \bibfield  {author} {\bibinfo {author} {\bibfnamefont {A.}~\bibnamefont
  {Micheli}}, \bibinfo {author} {\bibfnamefont {G.}~\bibnamefont {Pupillo}},
  \bibinfo {author} {\bibfnamefont {H.~P.}\ \bibnamefont {B\"uchler}}, \ and\
  \bibinfo {author} {\bibfnamefont {P.}~\bibnamefont {Zoller}},\ }\bibfield
  {title} {\enquote {\bibinfo {title} {Cold polar molecules in two-dimensional
  traps: Tailoring interactions with external fields for novel quantum
  phases},}\ }\href {\doibase 10.1103/PhysRevA.76.043604} {\bibfield  {journal}
  {\bibinfo  {journal} {Phys. Rev. A}\ }\textbf {\bibinfo {volume} {76}},\
  \bibinfo {pages} {043604} (\bibinfo {year} {2007})}\BibitemShut {NoStop}%
\bibitem [{\citenamefont {Andr\'e}\ \emph {et~al.}(2006)\citenamefont
  {Andr\'e}, \citenamefont {DeMille}, \citenamefont {Doyle}, \citenamefont
  {Lukin}, \citenamefont {Maxwell}, \citenamefont {Rabl}, \citenamefont
  {Schoelkopf},\ and\ \citenamefont {Zoller}}]{Andre2006}%
  \BibitemOpen
  \bibfield  {author} {\bibinfo {author} {\bibfnamefont {A.}~\bibnamefont
  {Andr\'e}}, \bibinfo {author} {\bibfnamefont {D.}~\bibnamefont {DeMille}},
  \bibinfo {author} {\bibfnamefont {J.~M.}\ \bibnamefont {Doyle}}, \bibinfo
  {author} {\bibfnamefont {M.~D.}\ \bibnamefont {Lukin}}, \bibinfo {author}
  {\bibfnamefont {S.~E.}\ \bibnamefont {Maxwell}}, \bibinfo {author}
  {\bibfnamefont {P.}~\bibnamefont {Rabl}}, \bibinfo {author} {\bibfnamefont
  {R.~J.}\ \bibnamefont {Schoelkopf}}, \ and\ \bibinfo {author} {\bibfnamefont
  {P.}~\bibnamefont {Zoller}},\ }\bibfield  {title} {\enquote {\bibinfo {title}
  {A coherent all-electrical interface between polar molecules and mesoscopic
  superconducting resonators},}\ }\href {\doibase 10.1038/nphys386} {\bibfield
  {journal} {\bibinfo  {journal} {Nat. Phys.}\ }\textbf {\bibinfo {volume}
  {2}},\ \bibinfo {pages} {636--642} (\bibinfo {year} {2006})}\BibitemShut
  {NoStop}%
\bibitem [{\citenamefont {Rabl}\ \emph {et~al.}(2006)\citenamefont {Rabl},
  \citenamefont {DeMille}, \citenamefont {Doyle}, \citenamefont {Lukin},
  \citenamefont {Schoelkopf},\ and\ \citenamefont {Zoller}}]{Rabl2006}%
  \BibitemOpen
  \bibfield  {author} {\bibinfo {author} {\bibfnamefont {P.}~\bibnamefont
  {Rabl}}, \bibinfo {author} {\bibfnamefont {D.}~\bibnamefont {DeMille}},
  \bibinfo {author} {\bibfnamefont {J.~M.}\ \bibnamefont {Doyle}}, \bibinfo
  {author} {\bibfnamefont {M.~D.}\ \bibnamefont {Lukin}}, \bibinfo {author}
  {\bibfnamefont {R.~J.}\ \bibnamefont {Schoelkopf}}, \ and\ \bibinfo {author}
  {\bibfnamefont {P.}~\bibnamefont {Zoller}},\ }\bibfield  {title} {\enquote
  {\bibinfo {title} {{Hybrid Quantum Processors: Molecular Ensembles as Quantum
  Memory for Solid State Circuits}},}\ }\href {\doibase
  10.1103/PhysRevLett.97.033003} {\bibfield  {journal} {\bibinfo  {journal}
  {Phys. Rev. Lett.}\ }\textbf {\bibinfo {volume} {97}},\ \bibinfo {pages}
  {033003} (\bibinfo {year} {2006})}\BibitemShut {NoStop}%
\bibitem [{\citenamefont {DeMille}(2002)}]{DeMille2002}%
  \BibitemOpen
  \bibfield  {author} {\bibinfo {author} {\bibfnamefont {D.}~\bibnamefont
  {DeMille}},\ }\bibfield  {title} {\enquote {\bibinfo {title} {{Quantum
  Computation with Trapped Polar Molecules}},}\ }\href {\doibase
  10.1103/PhysRevLett.88.067901} {\bibfield  {journal} {\bibinfo  {journal}
  {Phys. Rev. Lett.}\ }\textbf {\bibinfo {volume} {88}},\ \bibinfo {pages}
  {067901} (\bibinfo {year} {2002})}\BibitemShut {NoStop}%
\bibitem [{\citenamefont {Peter}\ \emph {et~al.}(2012)\citenamefont {Peter},
  \citenamefont {M{\"{u}}ller}, \citenamefont {Wessel},\ and\ \citenamefont
  {B{\"{u}}chler}}]{Peter2012}%
  \BibitemOpen
  \bibfield  {author} {\bibinfo {author} {\bibfnamefont {D.}~\bibnamefont
  {Peter}}, \bibinfo {author} {\bibfnamefont {S.}~\bibnamefont {M{\"{u}}ller}},
  \bibinfo {author} {\bibfnamefont {S.}~\bibnamefont {Wessel}}, \ and\ \bibinfo
  {author} {\bibfnamefont {H.~P.}\ \bibnamefont {B{\"{u}}chler}},\ }\bibfield
  {title} {\enquote {\bibinfo {title} {{Anomalous Behavior of Spin Systems with
  Dipolar Interactions}},}\ }\href {\doibase 10.1103/PhysRevLett.109.025303}
  {\bibfield  {journal} {\bibinfo  {journal} {Phys. Rev. Lett.}\ }\textbf
  {\bibinfo {volume} {109}},\ \bibinfo {pages} {025303} (\bibinfo {year}
  {2012})}\BibitemShut {NoStop}%
\bibitem [{\citenamefont {Yao}\ \emph {et~al.}(2013)\citenamefont {Yao},
  \citenamefont {Gorshkov}, \citenamefont {Laumann}, \citenamefont
  {L{\"{a}}uchli}, \citenamefont {Ye},\ and\ \citenamefont {Lukin}}]{Yao2013}%
  \BibitemOpen
  \bibfield  {author} {\bibinfo {author} {\bibfnamefont {N.~Y.}\ \bibnamefont
  {Yao}}, \bibinfo {author} {\bibfnamefont {A.~V.}\ \bibnamefont {Gorshkov}},
  \bibinfo {author} {\bibfnamefont {C.~R.}\ \bibnamefont {Laumann}}, \bibinfo
  {author} {\bibfnamefont {A.~M.}\ \bibnamefont {L{\"{a}}uchli}}, \bibinfo
  {author} {\bibfnamefont {J.}~\bibnamefont {Ye}}, \ and\ \bibinfo {author}
  {\bibfnamefont {M.~D.}\ \bibnamefont {Lukin}},\ }\bibfield  {title} {\enquote
  {\bibinfo {title} {{Realizing Fractional Chern Insulators in Dipolar Spin
  Systems}},}\ }\href {\doibase 10.1103/PhysRevLett.110.185302} {\bibfield
  {journal} {\bibinfo  {journal} {Phys. Rev. Lett.}\ }\textbf {\bibinfo
  {volume} {110}},\ \bibinfo {pages} {185302} (\bibinfo {year}
  {2013})}\BibitemShut {NoStop}%
\bibitem [{\citenamefont {Micheli}\ \emph {et~al.}(2006)\citenamefont
  {Micheli}, \citenamefont {Brennen},\ and\ \citenamefont
  {Zoller}}]{Micheli2006}%
  \BibitemOpen
  \bibfield  {author} {\bibinfo {author} {\bibfnamefont {A.}~\bibnamefont
  {Micheli}}, \bibinfo {author} {\bibfnamefont {G.~K.}\ \bibnamefont
  {Brennen}}, \ and\ \bibinfo {author} {\bibfnamefont {P.}~\bibnamefont
  {Zoller}},\ }\bibfield  {title} {\enquote {\bibinfo {title} {{A toolbox for
  lattice-spin models with polar molecules}},}\ }\href
  {https://doi.org/10.1038/nphys287} {\bibfield  {journal} {\bibinfo  {journal}
  {Nat. Phys.}\ }\textbf {\bibinfo {volume} {2}},\ \bibinfo {pages} {341--347}
  (\bibinfo {year} {2006})}\BibitemShut {NoStop}%
\bibitem [{\citenamefont {Karman}\ and\ \citenamefont
  {Hutson}(2018)}]{Karman2018}%
  \BibitemOpen
  \bibfield  {author} {\bibinfo {author} {\bibfnamefont {T.}~\bibnamefont
  {Karman}}\ and\ \bibinfo {author} {\bibfnamefont {J.~M.}\ \bibnamefont
  {Hutson}},\ }\bibfield  {title} {\enquote {\bibinfo {title} {Microwave
  shielding of ultracold polar molecules},}\ }\href {\doibase
  10.1103/PhysRevLett.121.163401} {\bibfield  {journal} {\bibinfo  {journal}
  {Phys. Rev. Lett.}\ }\textbf {\bibinfo {volume} {121}},\ \bibinfo {pages}
  {163401} (\bibinfo {year} {2018})}\BibitemShut {NoStop}%
\bibitem [{\citenamefont {Park}\ \emph {et~al.}(2017)\citenamefont {Park},
  \citenamefont {Yan}, \citenamefont {Loh}, \citenamefont {Will},\ and\
  \citenamefont {Zwierlein}}]{Park2017}%
  \BibitemOpen
  \bibfield  {author} {\bibinfo {author} {\bibfnamefont {J.~W.}\ \bibnamefont
  {Park}}, \bibinfo {author} {\bibfnamefont {Z.~Z.}\ \bibnamefont {Yan}},
  \bibinfo {author} {\bibfnamefont {H.}~\bibnamefont {Loh}}, \bibinfo {author}
  {\bibfnamefont {S.~A.}\ \bibnamefont {Will}}, \ and\ \bibinfo {author}
  {\bibfnamefont {M.~W.}\ \bibnamefont {Zwierlein}},\ }\bibfield  {title}
  {\enquote {\bibinfo {title} {{Second-scale nuclear spin coherence time of
  ultracold $^{23}$Na$^{40}$K molecules}},}\ }\href {\doibase
  10.1126/SCIENCE.AAL5066} {\bibfield  {journal} {\bibinfo  {journal}
  {Science}\ }\textbf {\bibinfo {volume} {357}},\ \bibinfo {pages} {372--375}
  (\bibinfo {year} {2017})}\BibitemShut {NoStop}%
\bibitem [{\citenamefont {Kotochigova}\ and\ \citenamefont
  {DeMille}(2010)}]{Kotochigova2010}%
  \BibitemOpen
  \bibfield  {author} {\bibinfo {author} {\bibfnamefont {S.}~\bibnamefont
  {Kotochigova}}\ and\ \bibinfo {author} {\bibfnamefont {D.}~\bibnamefont
  {DeMille}},\ }\bibfield  {title} {\enquote {\bibinfo {title}
  {Electric-field-dependent dynamic polarizability and state-insensitive
  conditions for optical trapping of diatomic polar molecules},}\ }\href
  {\doibase 10.1103/PhysRevA.82.063421} {\bibfield  {journal} {\bibinfo
  {journal} {Phys. Rev. A}\ }\textbf {\bibinfo {volume} {82}},\ \bibinfo
  {pages} {063421} (\bibinfo {year} {2010})}\BibitemShut {NoStop}%
\bibitem [{\citenamefont {See\ss{}elberg}\ \emph {et~al.}(2018)\citenamefont
  {See\ss{}elberg}, \citenamefont {Luo}, \citenamefont {Li}, \citenamefont
  {Bause}, \citenamefont {Kotochigova}, \citenamefont {Bloch},\ and\
  \citenamefont {Gohle}}]{Gohle2018}%
  \BibitemOpen
  \bibfield  {author} {\bibinfo {author} {\bibfnamefont {F.}~\bibnamefont
  {See\ss{}elberg}}, \bibinfo {author} {\bibfnamefont {X.-Y.}\ \bibnamefont
  {Luo}}, \bibinfo {author} {\bibfnamefont {M.}~\bibnamefont {Li}}, \bibinfo
  {author} {\bibfnamefont {R.}~\bibnamefont {Bause}}, \bibinfo {author}
  {\bibfnamefont {S.}~\bibnamefont {Kotochigova}}, \bibinfo {author}
  {\bibfnamefont {I.}~\bibnamefont {Bloch}}, \ and\ \bibinfo {author}
  {\bibfnamefont {C.}~\bibnamefont {Gohle}},\ }\bibfield  {title} {\enquote
  {\bibinfo {title} {{Extending Rotational Coherence of Interacting Polar
  Molecules in a Spin-Decoupled Magic Trap}},}\ }\href {\doibase
  10.1103/PhysRevLett.121.253401} {\bibfield  {journal} {\bibinfo  {journal}
  {Phys. Rev. Lett.}\ }\textbf {\bibinfo {volume} {121}},\ \bibinfo {pages}
  {253401} (\bibinfo {year} {2018})}\BibitemShut {NoStop}%
\bibitem [{\citenamefont {Kondov}\ \emph {et~al.}(2019)\citenamefont {Kondov},
  \citenamefont {Lee}, \citenamefont {Leung}, \citenamefont {Liedl},
  \citenamefont {Majewska}, \citenamefont {Moszynski},\ and\ \citenamefont
  {Zelevinsky}}]{Kondov2019}%
  \BibitemOpen
  \bibfield  {author} {\bibinfo {author} {\bibfnamefont {S.~S.}\ \bibnamefont
  {Kondov}}, \bibinfo {author} {\bibfnamefont {C.-H.}\ \bibnamefont {Lee}},
  \bibinfo {author} {\bibfnamefont {K.~H.}\ \bibnamefont {Leung}}, \bibinfo
  {author} {\bibfnamefont {C.}~\bibnamefont {Liedl}}, \bibinfo {author}
  {\bibfnamefont {I.}~\bibnamefont {Majewska}}, \bibinfo {author}
  {\bibfnamefont {R.}~\bibnamefont {Moszynski}}, \ and\ \bibinfo {author}
  {\bibfnamefont {T.}~\bibnamefont {Zelevinsky}},\ }\bibfield  {title}
  {\enquote {\bibinfo {title} {{Molecular lattice clock with long vibrational
  coherence}},}\ }\href {\doibase 10.1038/s41567-019-0632-3} {\bibfield
  {journal} {\bibinfo  {journal} {Nat. Phys.}\ }\textbf {\bibinfo {volume}
  {15}},\ \bibinfo {pages} {1118–1122} (\bibinfo {year} {2019})}\BibitemShut
  {NoStop}%
\bibitem [{\citenamefont {Barry}\ \emph {et~al.}(2014)\citenamefont {Barry},
  \citenamefont {McCarron}, \citenamefont {Norrgard}, \citenamefont
  {Steinecker},\ and\ \citenamefont {DeMille}}]{Barry2014}%
  \BibitemOpen
  \bibfield  {author} {\bibinfo {author} {\bibfnamefont {J.~F.}\ \bibnamefont
  {Barry}}, \bibinfo {author} {\bibfnamefont {D.~J.}\ \bibnamefont {McCarron}},
  \bibinfo {author} {\bibfnamefont {E.~B.}\ \bibnamefont {Norrgard}}, \bibinfo
  {author} {\bibfnamefont {M.~H.}\ \bibnamefont {Steinecker}}, \ and\ \bibinfo
  {author} {\bibfnamefont {D.}~\bibnamefont {DeMille}},\ }\bibfield  {title}
  {\enquote {\bibinfo {title} {Magneto-optical trapping of a diatomic
  molecule},}\ }\href {\doibase 10.1038/nature13634} {\bibfield  {journal}
  {\bibinfo  {journal} {Nature}\ }\textbf {\bibinfo {volume} {512}},\ \bibinfo
  {pages} {286--289} (\bibinfo {year} {2014})}\BibitemShut {NoStop}%
\bibitem [{\citenamefont {Williams}\ \emph {et~al.}(2017)\citenamefont
  {Williams}, \citenamefont {Truppe}, \citenamefont {Hambach}, \citenamefont
  {Caldwell}, \citenamefont {Fitch}, \citenamefont {Hinds}, \citenamefont
  {Sauer},\ and\ \citenamefont {Tarbutt}}]{Williams_2017}%
  \BibitemOpen
  \bibfield  {author} {\bibinfo {author} {\bibfnamefont {H.~J.}\ \bibnamefont
  {Williams}}, \bibinfo {author} {\bibfnamefont {S.}~\bibnamefont {Truppe}},
  \bibinfo {author} {\bibfnamefont {M.}~\bibnamefont {Hambach}}, \bibinfo
  {author} {\bibfnamefont {L.}~\bibnamefont {Caldwell}}, \bibinfo {author}
  {\bibfnamefont {N.~J.}\ \bibnamefont {Fitch}}, \bibinfo {author}
  {\bibfnamefont {E.~A.}\ \bibnamefont {Hinds}}, \bibinfo {author}
  {\bibfnamefont {B.~E.}\ \bibnamefont {Sauer}}, \ and\ \bibinfo {author}
  {\bibfnamefont {M.~R.}\ \bibnamefont {Tarbutt}},\ }\bibfield  {title}
  {\enquote {\bibinfo {title} {Characteristics of a magneto-optical trap of
  molecules},}\ }\href {\doibase 10.1088/1367-2630/aa8e52} {\bibfield
  {journal} {\bibinfo  {journal} {New J. Phys.}\ }\textbf {\bibinfo {volume}
  {19}},\ \bibinfo {pages} {113035} (\bibinfo {year} {2017})}\BibitemShut
  {NoStop}%
\bibitem [{\citenamefont {Anderegg}\ \emph {et~al.}(2017)\citenamefont
  {Anderegg}, \citenamefont {Augenbraun}, \citenamefont {Chae}, \citenamefont
  {Hemmerling}, \citenamefont {Hutzler}, \citenamefont {Ravi}, \citenamefont
  {Collopy}, \citenamefont {Ye}, \citenamefont {Ketterle},\ and\ \citenamefont
  {Doyle}}]{Anderegg2017}%
  \BibitemOpen
  \bibfield  {author} {\bibinfo {author} {\bibfnamefont {L.}~\bibnamefont
  {Anderegg}}, \bibinfo {author} {\bibfnamefont {B.~L.}\ \bibnamefont
  {Augenbraun}}, \bibinfo {author} {\bibfnamefont {E.}~\bibnamefont {Chae}},
  \bibinfo {author} {\bibfnamefont {B.}~\bibnamefont {Hemmerling}}, \bibinfo
  {author} {\bibfnamefont {N.~R.}\ \bibnamefont {Hutzler}}, \bibinfo {author}
  {\bibfnamefont {A.}~\bibnamefont {Ravi}}, \bibinfo {author} {\bibfnamefont
  {Al.}\ \bibnamefont {Collopy}}, \bibinfo {author} {\bibfnamefont
  {J.}~\bibnamefont {Ye}}, \bibinfo {author} {\bibfnamefont {W.}~\bibnamefont
  {Ketterle}}, \ and\ \bibinfo {author} {\bibfnamefont {J.~M.}\ \bibnamefont
  {Doyle}},\ }\bibfield  {title} {\enquote {\bibinfo {title} {{Radio Frequency
  Magneto-Optical Trapping of CaF with High Density}},}\ }\href {\doibase
  10.1103/PhysRevLett.119.103201} {\bibfield  {journal} {\bibinfo  {journal}
  {Phys. Rev. Lett.}\ }\textbf {\bibinfo {volume} {119}},\ \bibinfo {pages}
  {103201} (\bibinfo {year} {2017})}\BibitemShut {NoStop}%
\bibitem [{\citenamefont {Collopy}\ \emph {et~al.}(2018)\citenamefont
  {Collopy}, \citenamefont {Ding}, \citenamefont {Wu}, \citenamefont
  {Finneran}, \citenamefont {Anderegg}, \citenamefont {Augenbraun},
  \citenamefont {Doyle},\ and\ \citenamefont {Ye}}]{Collopy2018}%
  \BibitemOpen
  \bibfield  {author} {\bibinfo {author} {\bibfnamefont {A.~L.}\ \bibnamefont
  {Collopy}}, \bibinfo {author} {\bibfnamefont {S.}~\bibnamefont {Ding}},
  \bibinfo {author} {\bibfnamefont {Y.}~\bibnamefont {Wu}}, \bibinfo {author}
  {\bibfnamefont {I.~A.}\ \bibnamefont {Finneran}}, \bibinfo {author}
  {\bibfnamefont {L.}~\bibnamefont {Anderegg}}, \bibinfo {author}
  {\bibfnamefont {B.~L.}\ \bibnamefont {Augenbraun}}, \bibinfo {author}
  {\bibfnamefont {J.~M.}\ \bibnamefont {Doyle}}, \ and\ \bibinfo {author}
  {\bibfnamefont {J.}~\bibnamefont {Ye}},\ }\bibfield  {title} {\enquote
  {\bibinfo {title} {{3D Magneto-Optical Trap of Yttrium Monoxide}},}\ }\href
  {\doibase 10.1103/PhysRevLett.121.213201} {\bibfield  {journal} {\bibinfo
  {journal} {Phys. Rev. Lett.}\ }\textbf {\bibinfo {volume} {121}},\ \bibinfo
  {pages} {213201} (\bibinfo {year} {2018})}\BibitemShut {NoStop}%
\bibitem [{\citenamefont {McCarron}\ \emph {et~al.}(2018)\citenamefont
  {McCarron}, \citenamefont {Steinecker}, \citenamefont {Zhu},\ and\
  \citenamefont {DeMille}}]{McCarron2018}%
  \BibitemOpen
  \bibfield  {author} {\bibinfo {author} {\bibfnamefont {D.~J.}\ \bibnamefont
  {McCarron}}, \bibinfo {author} {\bibfnamefont {M.~H.}\ \bibnamefont
  {Steinecker}}, \bibinfo {author} {\bibfnamefont {Y.}~\bibnamefont {Zhu}}, \
  and\ \bibinfo {author} {\bibfnamefont {D.}~\bibnamefont {DeMille}},\
  }\bibfield  {title} {\enquote {\bibinfo {title} {Magnetic trapping of an
  ultracold gas of polar molecules},}\ }\href {\doibase
  10.1103/PhysRevLett.121.013202} {\bibfield  {journal} {\bibinfo  {journal}
  {Phys. Rev. Lett.}\ }\textbf {\bibinfo {volume} {121}},\ \bibinfo {pages}
  {013202} (\bibinfo {year} {2018})}\BibitemShut {NoStop}%
\bibitem [{\citenamefont {Williams}\ \emph {et~al.}(2018)\citenamefont
  {Williams}, \citenamefont {Caldwell}, \citenamefont {Fitch}, \citenamefont
  {Truppe}, \citenamefont {Rodewald}, \citenamefont {Hinds}, \citenamefont
  {Sauer},\ and\ \citenamefont {Tarbutt}}]{Williams2018}%
  \BibitemOpen
  \bibfield  {author} {\bibinfo {author} {\bibfnamefont {H.~J.}\ \bibnamefont
  {Williams}}, \bibinfo {author} {\bibfnamefont {L.}~\bibnamefont {Caldwell}},
  \bibinfo {author} {\bibfnamefont {N.~J.}\ \bibnamefont {Fitch}}, \bibinfo
  {author} {\bibfnamefont {S.}~\bibnamefont {Truppe}}, \bibinfo {author}
  {\bibfnamefont {J.}~\bibnamefont {Rodewald}}, \bibinfo {author}
  {\bibfnamefont {E.~A.}\ \bibnamefont {Hinds}}, \bibinfo {author}
  {\bibfnamefont {B.~E.}\ \bibnamefont {Sauer}}, \ and\ \bibinfo {author}
  {\bibfnamefont {M.~R.}\ \bibnamefont {Tarbutt}},\ }\bibfield  {title}
  {\enquote {\bibinfo {title} {Magnetic trapping and coherent control of
  laser-cooled molecules},}\ }\href {\doibase 10.1103/PhysRevLett.120.163201}
  {\bibfield  {journal} {\bibinfo  {journal} {Phys. Rev. Lett.}\ }\textbf
  {\bibinfo {volume} {120}},\ \bibinfo {pages} {163201} (\bibinfo {year}
  {2018})}\BibitemShut {NoStop}%
\bibitem [{\citenamefont {Wang}\ \emph {et~al.}(2017)\citenamefont {Wang},
  \citenamefont {Tran}, \citenamefont {Surendran}, \citenamefont {Herrera},
  \citenamefont {Balcytis}, \citenamefont {Nissen}, \citenamefont {Albrecht},
  \citenamefont {Sidorov},\ and\ \citenamefont {Hannaford}}]{Wang2017}%
  \BibitemOpen
  \bibfield  {author} {\bibinfo {author} {\bibfnamefont {Y.}~\bibnamefont
  {Wang}}, \bibinfo {author} {\bibfnamefont {T.}~\bibnamefont {Tran}}, \bibinfo
  {author} {\bibfnamefont {P.}~\bibnamefont {Surendran}}, \bibinfo {author}
  {\bibfnamefont {I.}~\bibnamefont {Herrera}}, \bibinfo {author} {\bibfnamefont
  {A.}~\bibnamefont {Balcytis}}, \bibinfo {author} {\bibfnamefont
  {D.}~\bibnamefont {Nissen}}, \bibinfo {author} {\bibfnamefont
  {M.}~\bibnamefont {Albrecht}}, \bibinfo {author} {\bibfnamefont
  {A.}~\bibnamefont {Sidorov}}, \ and\ \bibinfo {author} {\bibfnamefont
  {P.}~\bibnamefont {Hannaford}},\ }\bibfield  {title} {\enquote {\bibinfo
  {title} {{Trapping ultracold atoms in a sub-micron-period triangular magnetic
  lattice}},}\ }\href {\doibase 10.1103/PhysRevA.96.013630} {\bibfield
  {journal} {\bibinfo  {journal} {Phys. Rev. A}\ }\textbf {\bibinfo {volume}
  {96}},\ \bibinfo {pages} {013630} (\bibinfo {year} {2017})}\BibitemShut
  {NoStop}%
\bibitem [{\citenamefont {Brown}\ and\ \citenamefont
  {Carrington}(2003)}]{Brown2003}%
  \BibitemOpen
  \bibfield  {author} {\bibinfo {author} {\bibfnamefont {J.}~\bibnamefont
  {Brown}}\ and\ \bibinfo {author} {\bibfnamefont {A.}~\bibnamefont
  {Carrington}},\ }\href@noop {} {\emph {\bibinfo {title} {{Rotational
  Spectroscopy of Diatomic Molecules}}}}\ (\bibinfo  {publisher} {Cambridge
  University Press},\ \bibinfo {address} {Cambridge},\ \bibinfo {year}
  {2003})\BibitemShut {NoStop}%
\bibitem [{\citenamefont {Bowater}\ \emph {et~al.}(1973)\citenamefont
  {Bowater}, \citenamefont {Brown},\ and\ \citenamefont
  {Carrington}}]{Bowater1973}%
  \BibitemOpen
  \bibfield  {author} {\bibinfo {author} {\bibfnamefont {I.~C.}\ \bibnamefont
  {Bowater}}, \bibinfo {author} {\bibfnamefont {J.~M.}\ \bibnamefont {Brown}},
  \ and\ \bibinfo {author} {\bibfnamefont {A.}~\bibnamefont {Carrington}},\
  }\bibfield  {title} {\enquote {\bibinfo {title} {{Microwave Spectroscopy of
  Nonlinear Free Radicals. I. General Theory and Application to the Zeeman
  Effect in HCO}},}\ }\href {\doibase 10.1098/rspa.1973.0062} {\bibfield
  {journal} {\bibinfo  {journal} {Proc. R. Soc. A}\ }\textbf {\bibinfo {volume}
  {333}},\ \bibinfo {pages} {265--288} (\bibinfo {year} {1973})}\BibitemShut
  {NoStop}%
\bibitem [{\citenamefont {Neese}(2012)}]{Neese:2012}%
  \BibitemOpen
  \bibfield  {author} {\bibinfo {author} {\bibfnamefont {F.}~\bibnamefont
  {Neese}},\ }\bibfield  {title} {\enquote {\bibinfo {title} {The {ORCA}
  program system},}\ }\href {https://doi.org/10.1002/wcms.81} {\bibfield
  {journal} {\bibinfo  {journal} {Interdisc. Rev.: Comput. Mol. Sci.}\ }\textbf
  {\bibinfo {volume} {2}},\ \bibinfo {pages} {73--78} (\bibinfo {year}
  {2012})}\BibitemShut {NoStop}%
\bibitem [{\citenamefont {Stephens}\ \emph {et~al.}(1994)\citenamefont
  {Stephens}, \citenamefont {Devlin}, \citenamefont {Chabalowski},\ and\
  \citenamefont {Frisch}}]{Stephens:1994}%
  \BibitemOpen
  \bibfield  {author} {\bibinfo {author} {\bibfnamefont {P.~J.}\ \bibnamefont
  {Stephens}}, \bibinfo {author} {\bibfnamefont {F.~J.}\ \bibnamefont
  {Devlin}}, \bibinfo {author} {\bibfnamefont {C.~F.}\ \bibnamefont
  {Chabalowski}}, \ and\ \bibinfo {author} {\bibfnamefont {M.~J.}\ \bibnamefont
  {Frisch}},\ }\bibfield  {title} {\enquote {\bibinfo {title} {{Ab Initio
  Calculation of Vibrational Absorption and Circular Dichroism Spectra Using
  Density Functional Force Fields}},}\ }\href
  {https://doi.org/10.1021/j100096a001} {\bibfield  {journal} {\bibinfo
  {journal} {J. Phys. Chem.}\ }\textbf {\bibinfo {volume} {98}},\ \bibinfo
  {pages} {11623--11627} (\bibinfo {year} {1994})}\BibitemShut {NoStop}%
\bibitem [{\citenamefont {Pollak}\ and\ \citenamefont
  {Weigend}(2017)}]{Pollak:2017}%
  \BibitemOpen
  \bibfield  {author} {\bibinfo {author} {\bibfnamefont {P.}~\bibnamefont
  {Pollak}}\ and\ \bibinfo {author} {\bibfnamefont {F.}~\bibnamefont
  {Weigend}},\ }\bibfield  {title} {\enquote {\bibinfo {title} {Segmented
  contracted error-consistent basis sets of double- and triple-$\zeta$ valence
  quality for one- and two-component relativistic all-electron calculations},}\
  }\href {https://doi.org/10.1021/acs.jctc.7b00593} {\bibfield  {journal}
  {\bibinfo  {journal} {J. Chem. Theor. Comput.}\ }\textbf {\bibinfo {volume}
  {13}},\ \bibinfo {pages} {3696--3705} (\bibinfo {year} {2017})}\BibitemShut
  {NoStop}%
\bibitem [{\citenamefont {van Lenthe}\ \emph {et~al.}(1993)\citenamefont {van
  Lenthe}, \citenamefont {Baerends},\ and\ \citenamefont
  {Snijders}}]{vanLenthe:1993}%
  \BibitemOpen
  \bibfield  {author} {\bibinfo {author} {\bibfnamefont {E.}~\bibnamefont {van
  Lenthe}}, \bibinfo {author} {\bibfnamefont {E.~J.}\ \bibnamefont {Baerends}},
  \ and\ \bibinfo {author} {\bibfnamefont {J.~G.}\ \bibnamefont {Snijders}},\
  }\bibfield  {title} {\enquote {\bibinfo {title} {{Relativistic regular
  two-component Hamiltonians}},}\ }\href {https://doi.org/10.1063/1.466059}
  {\bibfield  {journal} {\bibinfo  {journal} {J. Comput. Phys.}\ }\textbf
  {\bibinfo {volume} {99}},\ \bibinfo {pages} {4597} (\bibinfo {year}
  {1993})}\BibitemShut {NoStop}%
\bibitem [{\citenamefont {van Lenthe}\ \emph {et~al.}(1994)\citenamefont {van
  Lenthe}, \citenamefont {Baerends},\ and\ \citenamefont
  {Snijders}}]{vanLenthe:1994}%
  \BibitemOpen
  \bibfield  {author} {\bibinfo {author} {\bibfnamefont {E.}~\bibnamefont {van
  Lenthe}}, \bibinfo {author} {\bibfnamefont {E.~J.}\ \bibnamefont {Baerends}},
  \ and\ \bibinfo {author} {\bibfnamefont {J.~G.}\ \bibnamefont {Snijders}},\
  }\bibfield  {title} {\enquote {\bibinfo {title} {Relativistic total energy
  using regular approximations},}\ }\href {\doibase 10.1063/1.467943}
  {\bibfield  {journal} {\bibinfo  {journal} {J. Comput. Phys.}\ }\textbf
  {\bibinfo {volume} {101}},\ \bibinfo {pages} {9783--9792} (\bibinfo {year}
  {1994})}\BibitemShut {NoStop}%
\bibitem [{\citenamefont {van Lenthe}\ \emph {et~al.}(1999)\citenamefont {van
  Lenthe}, \citenamefont {Baerends},\ and\ \citenamefont
  {Snijders}}]{vanLenthe:1999}%
  \BibitemOpen
  \bibfield  {author} {\bibinfo {author} {\bibfnamefont {E.}~\bibnamefont {van
  Lenthe}}, \bibinfo {author} {\bibfnamefont {E.~J.}\ \bibnamefont {Baerends}},
  \ and\ \bibinfo {author} {\bibfnamefont {J.~G.}\ \bibnamefont {Snijders}},\
  }\bibfield  {title} {\enquote {\bibinfo {title} {Geometry optimizations in
  the zero order regular approximation for relativistic effects},}\ }\href
  {https://doi.org/10.1063/1.478813} {\bibfield  {journal} {\bibinfo  {journal}
  {J. Comput. Phys.}\ }\textbf {\bibinfo {volume} {110}},\ \bibinfo {pages}
  {8943} (\bibinfo {year} {1999})}\BibitemShut {NoStop}%
\bibitem [{\citenamefont {Aidas}\ \emph {et~al.}(2014)\citenamefont {Aidas},
  \citenamefont {Angeli}, \citenamefont {Bak}, \citenamefont {Bakken},
  \citenamefont {Bast}, \citenamefont {Boman}, \citenamefont {Christiansen},
  \citenamefont {Cimiraglia}, \citenamefont {Coriani}, \citenamefont {Dahle},
  \citenamefont {Dalskov}, \citenamefont {Ekstr\"om},\ and\ \citenamefont
  {Enevoldsen}}]{Dalton:current}%
  \BibitemOpen
  \bibfield  {author} {\bibinfo {author} {\bibfnamefont {K.}~\bibnamefont
  {Aidas}}, \bibinfo {author} {\bibfnamefont {C.}~\bibnamefont {Angeli}},
  \bibinfo {author} {\bibfnamefont {K.~L.}\ \bibnamefont {Bak}}, \bibinfo
  {author} {\bibfnamefont {V.}~\bibnamefont {Bakken}}, \bibinfo {author}
  {\bibfnamefont {R.}~\bibnamefont {Bast}}, \bibinfo {author} {\bibfnamefont
  {L.}~\bibnamefont {Boman}}, \bibinfo {author} {\bibfnamefont
  {O.}~\bibnamefont {Christiansen}}, \bibinfo {author} {\bibfnamefont
  {R.}~\bibnamefont {Cimiraglia}}, \bibinfo {author} {\bibfnamefont
  {S.}~\bibnamefont {Coriani}}, \bibinfo {author} {\bibfnamefont
  {P.}~\bibnamefont {Dahle}}, \bibinfo {author} {\bibfnamefont {E.~K.}\
  \bibnamefont {Dalskov}}, \bibinfo {author} {\bibfnamefont {U.}~\bibnamefont
  {Ekstr\"om}}, \ and\ \bibinfo {author} {\bibfnamefont {T.}~\bibnamefont
  {Enevoldsen}},\ }\bibfield  {title} {\enquote {\bibinfo {title} {{DALTON}, a
  molecular electronic structure program, {R}elease 2017.alpha},}\ }\href@noop
  {} {\bibfield  {journal} {\bibinfo  {journal} {Wiley Interdisc Rev. Comput.
  Mol. Sci.}\ }\textbf {\bibinfo {volume} {4}},\ \bibinfo {pages} {269--284}
  (\bibinfo {year} {2014})},\ \bibinfo {note}
  {\url{http://daltonprogram.org}}\BibitemShut {NoStop}%
\bibitem [{Sup()}]{Supp}%
  \BibitemOpen
  \href@noop {} {}\bibinfo {note} {{See Supplemental Material for further
  details of electronic structure calculations of $g_\textrm{r}$ and $g_l$,
  which includes Refs.~[33-47]}}\BibitemShut {NoStop}%
\bibitem [{\citenamefont {Aldegunde}\ and\ \citenamefont
  {Hutson}(2018)}]{aldegunde:2Sigma:pra2019}%
  \BibitemOpen
  \bibfield  {author} {\bibinfo {author} {\bibfnamefont {J.}~\bibnamefont
  {Aldegunde}}\ and\ \bibinfo {author} {\bibfnamefont {J.~M.}\ \bibnamefont
  {Hutson}},\ }\bibfield  {title} {\enquote {\bibinfo {title} {Hyperfine
  structure of {$^2\Sigma$} molecules containing alkaline-earth-metal atoms},}\
  }\href {https://journals.aps.org/pra/abstract/10.1103/PhysRevA.97.042505}
  {\bibfield  {journal} {\bibinfo  {journal} {Phys. Rev. A}\ }\textbf {\bibinfo
  {volume} {97}},\ \bibinfo {pages} {042505} (\bibinfo {year}
  {2018})}\BibitemShut {NoStop}%
\bibitem [{\citenamefont {Curl}(1965)}]{curl:mp1965}%
  \BibitemOpen
  \bibfield  {author} {\bibinfo {author} {\bibfnamefont {R.~F.}\ \bibnamefont
  {Curl}},\ }\bibfield  {title} {\enquote {\bibinfo {title} {The relationship
  between electron spin rotation coupling constants and $g$-tensor
  components},}\ }\href {https://doi.org/10.1080/00268976500100761} {\bibfield
  {journal} {\bibinfo  {journal} {Mol. Phys.}\ }\textbf {\bibinfo {volume}
  {9}},\ \bibinfo {pages} {585} (\bibinfo {year} {1965})}\BibitemShut {NoStop}%
\bibitem [{\citenamefont {Anderson}\ \emph {et~al.}(1994)\citenamefont
  {Anderson}, \citenamefont {Allen},\ and\ \citenamefont
  {Ziurys}}]{Anderson:JCP1994:MgF}%
  \BibitemOpen
  \bibfield  {author} {\bibinfo {author} {\bibfnamefont {M.~A.}\ \bibnamefont
  {Anderson}}, \bibinfo {author} {\bibfnamefont {M.~D.}\ \bibnamefont {Allen}},
  \ and\ \bibinfo {author} {\bibfnamefont {L.~M.}\ \bibnamefont {Ziurys}},\
  }\bibfield  {title} {\enquote {\bibinfo {title} {{Millimeter‐wave
  spectroscopy of MgF: Structure and bonding in alkaline-earth monofluoride
  radicals}},}\ }\href {https://doi.org/10.1063/1.466565} {\bibfield  {journal}
  {\bibinfo  {journal} {J. Chem. Phys.}\ }\textbf {\bibinfo {volume} {100}},\
  \bibinfo {pages} {824--830} (\bibinfo {year} {1994})}\BibitemShut {NoStop}%
\bibitem [{\citenamefont {Childs}\ \emph
  {et~al.}(1981{\natexlab{a}})\citenamefont {Childs}, \citenamefont {Goodman},\
  and\ \citenamefont {Renhorn}}]{Childs:JMS1981:SrF}%
  \BibitemOpen
  \bibfield  {author} {\bibinfo {author} {\bibfnamefont {W.~J.}\ \bibnamefont
  {Childs}}, \bibinfo {author} {\bibfnamefont {L.~S.}\ \bibnamefont {Goodman}},
  \ and\ \bibinfo {author} {\bibfnamefont {I.}~\bibnamefont {Renhorn}},\
  }\bibfield  {title} {\enquote {\bibinfo {title} {{Radio-Frequency Optical
  Double-Resonance Spectrum of SrF: The $X^2\Sigma^+$ State}},}\ }\href
  {https://doi.org/10.1016/0022-2852(81)90422-7} {\bibfield  {journal}
  {\bibinfo  {journal} {J. Mol. Spectrosc.}\ }\textbf {\bibinfo {volume}
  {87}},\ \bibinfo {pages} {522--533} (\bibinfo {year}
  {1981}{\natexlab{a}})}\BibitemShut {NoStop}%
\bibitem [{\citenamefont {Knight}\ \emph {et~al.}(1972)\citenamefont {Knight},
  \citenamefont {Brom},\ and\ \citenamefont {Weltner}}]{Knight:JCP1972:BeH}%
  \BibitemOpen
  \bibfield  {author} {\bibinfo {author} {\bibfnamefont {L.~B.}\ \bibnamefont
  {Knight}, \bibfnamefont {Jr.}}, \bibinfo {author} {\bibfnamefont {J.~M.}\
  \bibnamefont {Brom}, \bibfnamefont {Jr.}}, \ and\ \bibinfo {author}
  {\bibfnamefont {W.}~\bibnamefont {Weltner}, \bibfnamefont {Jr.}},\ }\bibfield
   {title} {\enquote {\bibinfo {title} {{Hyperfine Interaction and Chemical
  Bonding in the BeH Molecule}},}\ }\href {https://doi.org/10.1063/1.1677336}
  {\bibfield  {journal} {\bibinfo  {journal} {J. Chem. Phys.}\ }\textbf
  {\bibinfo {volume} {56}},\ \bibinfo {pages} {1152--1155} (\bibinfo {year}
  {1972})}\BibitemShut {NoStop}%
\bibitem [{\citenamefont {Bernath}\ \emph {et~al.}(1985)\citenamefont
  {Bernath}, \citenamefont {Black},\ and\ \citenamefont
  {Brault}}]{Bernath:TAJ1985:MgH}%
  \BibitemOpen
  \bibfield  {author} {\bibinfo {author} {\bibfnamefont {P.~F.}\ \bibnamefont
  {Bernath}}, \bibinfo {author} {\bibfnamefont {J.~H.}\ \bibnamefont {Black}},
  \ and\ \bibinfo {author} {\bibfnamefont {J.~W.}\ \bibnamefont {Brault}},\
  }\bibfield  {title} {\enquote {\bibinfo {title} {The spectrum of magnesium
  hydride},}\ }\href {https://doi.org/10.1086/163620} {\bibfield  {journal}
  {\bibinfo  {journal} {Astrophys. J.}\ }\textbf {\bibinfo {volume} {298}},\
  \bibinfo {pages} {375--381} (\bibinfo {year} {1985})}\BibitemShut {NoStop}%
\bibitem [{\citenamefont {Berg}\ and\ \citenamefont
  {Klynning}(1974)}]{Berg1974:CaH}%
  \BibitemOpen
  \bibfield  {author} {\bibinfo {author} {\bibfnamefont {L.-E.}\ \bibnamefont
  {Berg}}\ and\ \bibinfo {author} {\bibfnamefont {L.}~\bibnamefont
  {Klynning}},\ }\bibfield  {title} {\enquote {\bibinfo {title} {{Rotational
  Analysis of the A-X and B-X Band Systems of CaH}},}\ }\href
  {https://doi.org/10.1088/0031-8949/10/6/009} {\bibfield  {journal} {\bibinfo
  {journal} {Phys. Scr.}\ }\textbf {\bibinfo {volume} {10}},\ \bibinfo {pages}
  {331--336} (\bibinfo {year} {1974})}\BibitemShut {NoStop}%
\bibitem [{\citenamefont {Knight}\ and\ \citenamefont {Weltner}(1971)}]{:SrH}%
  \BibitemOpen
  \bibfield  {author} {\bibinfo {author} {\bibfnamefont {L.~B.}\ \bibnamefont
  {Knight}, \bibfnamefont {Jr.}}\ and\ \bibinfo {author} {\bibfnamefont
  {W.}~\bibnamefont {Weltner}, \bibfnamefont {Jr.}},\ }\bibfield  {title}
  {\enquote {\bibinfo {title} {{Hyperfine Interaction and Chemical Bonding in
  MgH, CaH, SrH, and BaH Molecules}},}\ }\href
  {http://dx.doi.org/10.1063/1.1675441} {\bibfield  {journal} {\bibinfo
  {journal} {J. Chem. Phys.}\ }\textbf {\bibinfo {volume} {54}},\ \bibinfo
  {pages} {3875--3884} (\bibinfo {year} {1971})}\BibitemShut {NoStop}%
\bibitem [{\citenamefont {Brom}\ and\ \citenamefont
  {Weltner}(1976)}]{Brom:JCP1976:BeOH}%
  \BibitemOpen
  \bibfield  {author} {\bibinfo {author} {\bibfnamefont {J.~M.}\ \bibnamefont
  {Brom}, \bibfnamefont {Jr.}}\ and\ \bibinfo {author} {\bibfnamefont
  {W.}~\bibnamefont {Weltner}, \bibfnamefont {Jr.}},\ }\bibfield  {title}
  {\enquote {\bibinfo {title} {{ESR spectrum of the BeOH molecule}},}\ }\href
  {http://dx.doi.org/10.1063/1.432671} {\bibfield  {journal} {\bibinfo
  {journal} {J. Chem. Phys.}\ }\textbf {\bibinfo {volume} {64}},\ \bibinfo
  {pages} {3894--3895} (\bibinfo {year} {1976})}\BibitemShut {NoStop}%
\bibitem [{\citenamefont {Brom}\ and\ \citenamefont
  {Weltner}(1973)}]{Brom:JCP:1973:MgOH}%
  \BibitemOpen
  \bibfield  {author} {\bibinfo {author} {\bibfnamefont {J.~M.}\ \bibnamefont
  {Brom}, \bibfnamefont {Jr.}}\ and\ \bibinfo {author} {\bibfnamefont
  {W.}~\bibnamefont {Weltner}, \bibfnamefont {Jr.}},\ }\bibfield  {title}
  {\enquote {\bibinfo {title} {{ESR spectrum and structure of the MgOH
  radical}},}\ }\href {https://doi.org/10.1063/1.1679147} {\bibfield  {journal}
  {\bibinfo  {journal} {J. Chem. Phys.}\ }\textbf {\bibinfo {volume} {58}},\
  \bibinfo {pages} {5322--5330} (\bibinfo {year} {1973})}\BibitemShut {NoStop}%
\bibitem [{\citenamefont {Ziurys}\ \emph {et~al.}(1992)\citenamefont {Ziurys},
  \citenamefont {Barclay},\ and\ \citenamefont
  {Anderson}}]{Ziurys:TAJ1992:CaOH}%
  \BibitemOpen
  \bibfield  {author} {\bibinfo {author} {\bibfnamefont {L.~M.}\ \bibnamefont
  {Ziurys}}, \bibinfo {author} {\bibfnamefont {W.~L.}\ \bibnamefont {Barclay},
  \bibfnamefont {Jr.}}, \ and\ \bibinfo {author} {\bibfnamefont {M.~A.}\
  \bibnamefont {Anderson}},\ }\bibfield  {title} {\enquote {\bibinfo {title}
  {{The millimeter-wave spectrum of the CaOH radical ($X^2\Sigma^+$)}},}\
  }\href {https://doi.org/10.1086/186262} {\bibfield  {journal} {\bibinfo
  {journal} {Astrophys. J.}\ }\textbf {\bibinfo {volume} {384}},\ \bibinfo
  {pages} {L63--L66} (\bibinfo {year} {1992})}\BibitemShut {NoStop}%
\bibitem [{\citenamefont {Anderson}\ \emph {et~al.}(1992)\citenamefont
  {Anderson}, \citenamefont {Barclay},\ and\ \citenamefont
  {Ziurys}}]{Anderson:CPL1992:SrOH}%
  \BibitemOpen
  \bibfield  {author} {\bibinfo {author} {\bibfnamefont {M.~A.}\ \bibnamefont
  {Anderson}}, \bibinfo {author} {\bibfnamefont {W.~L.}\ \bibnamefont
  {Barclay}}, \ and\ \bibinfo {author} {\bibfnamefont {L.~M.}\ \bibnamefont
  {Ziurys}},\ }\bibfield  {title} {\enquote {\bibinfo {title} {{The
  millimeter-wave spectrum of the SrOH and SrOD radicals}},}\ }\href
  {https://doi.org/10.1016/0009-2614(92)85948-A} {\bibfield  {journal}
  {\bibinfo  {journal} {Chem. Phys. Lett.}\ }\textbf {\bibinfo {volume}
  {196}},\ \bibinfo {pages} {166--172} (\bibinfo {year} {1992})}\BibitemShut
  {NoStop}%
\bibitem [{\citenamefont {Weltner}(1983)}]{Weltner:book}%
  \BibitemOpen
  \bibfield  {author} {\bibinfo {author} {\bibfnamefont {W.}~\bibnamefont
  {Weltner}, \bibfnamefont {Jr.}},\ }\enquote {\bibinfo {title} {Magnetic atoms
  and molecules},}\ \ (\bibinfo  {publisher} {Dover},\ \bibinfo {address} {New
  York},\ \bibinfo {year} {1983})\ pp.\ \bibinfo {pages} {50--51}\BibitemShut
  {NoStop}%
\bibitem [{\citenamefont {Mohn}\ \emph {et~al.}(2005)\citenamefont {Mohn},
  \citenamefont {Wilson}, \citenamefont {Lutn{\ae}s}, \citenamefont
  {Helgaker},\ and\ \citenamefont {Ruud}}]{Mohn:AQC2005}%
  \BibitemOpen
  \bibfield  {author} {\bibinfo {author} {\bibfnamefont {C.~E.}\ \bibnamefont
  {Mohn}}, \bibinfo {author} {\bibfnamefont {D.~J.~D.}\ \bibnamefont {Wilson}},
  \bibinfo {author} {\bibfnamefont {O.~B.}\ \bibnamefont {Lutn{\ae}s}},
  \bibinfo {author} {\bibfnamefont {T.}~\bibnamefont {Helgaker}}, \ and\
  \bibinfo {author} {\bibfnamefont {K.}~\bibnamefont {Ruud}},\ }\bibfield
  {title} {\enquote {\bibinfo {title} {The rotational $g$ tensor as a benchmark
  for ab initio molecular property calculations},}\ }\href
  {https://doi.org/10.1016/S0065-3276(05)50005-4} {\bibfield  {journal}
  {\bibinfo  {journal} {Adv. Quantum Chem.}\ }\textbf {\bibinfo {volume}
  {50}},\ \bibinfo {pages} {77--90} (\bibinfo {year} {2005})}\BibitemShut
  {NoStop}%
\bibitem [{\citenamefont {Lee}\ and\ \citenamefont {Taylor}(1989)}]{Lee:T1}%
  \BibitemOpen
  \bibfield  {author} {\bibinfo {author} {\bibfnamefont {T.~J.}\ \bibnamefont
  {Lee}}\ and\ \bibinfo {author} {\bibfnamefont {P.~R.}\ \bibnamefont
  {Taylor}},\ }\bibfield  {title} {\enquote {\bibinfo {title} {A diagnostic for
  determining the quality of single-reference electron correlation methods},}\
  }\href {https://doi.org/10.1002/qua.560360824} {\bibfield  {journal}
  {\bibinfo  {journal} {Int. J. Quantum Chem.}\ }\textbf {\bibinfo {volume}
  {36}},\ \bibinfo {pages} {199--207} (\bibinfo {year} {1989})}\BibitemShut
  {NoStop}%
\bibitem [{\citenamefont {Truppe}\ \emph {et~al.}(2017)\citenamefont {Truppe},
  \citenamefont {Williams}, \citenamefont {Hambach}, \citenamefont {Caldwell},
  \citenamefont {Fitch}, \citenamefont {Hinds}, \citenamefont {Sauer},\ and\
  \citenamefont {Tarbutt}}]{Truppe2017b}%
  \BibitemOpen
  \bibfield  {author} {\bibinfo {author} {\bibfnamefont {S.}~\bibnamefont
  {Truppe}}, \bibinfo {author} {\bibfnamefont {H.~J.}\ \bibnamefont
  {Williams}}, \bibinfo {author} {\bibfnamefont {M.}~\bibnamefont {Hambach}},
  \bibinfo {author} {\bibfnamefont {L.}~\bibnamefont {Caldwell}}, \bibinfo
  {author} {\bibfnamefont {N.~J.}\ \bibnamefont {Fitch}}, \bibinfo {author}
  {\bibfnamefont {E.~A.}\ \bibnamefont {Hinds}}, \bibinfo {author}
  {\bibfnamefont {B.~E.}\ \bibnamefont {Sauer}}, \ and\ \bibinfo {author}
  {\bibfnamefont {M.~R.}\ \bibnamefont {Tarbutt}},\ }\bibfield  {title}
  {\enquote {\bibinfo {title} {Molecules cooled below the doppler limit},}\
  }\href {https://doi.org/10.1038/nphys4241} {\bibfield  {journal} {\bibinfo
  {journal} {Nat. Phys.}\ }\textbf {\bibinfo {volume} {13}} (\bibinfo {year}
  {2017})}\BibitemShut {NoStop}%
\bibitem [{\citenamefont {Blackmore}\ \emph {et~al.}(2018)\citenamefont
  {Blackmore}, \citenamefont {Caldwell}, \citenamefont {Gregory}, \citenamefont
  {Bridge}, \citenamefont {Sawant}, \citenamefont {Aldegunde}, \citenamefont
  {Mur-Petit}, \citenamefont {Jaksch}, \citenamefont {Hutson}, \citenamefont
  {Sauer}, \citenamefont {Tarbutt},\ and\ \citenamefont
  {Cornish}}]{Blackmore2018}%
  \BibitemOpen
  \bibfield  {author} {\bibinfo {author} {\bibfnamefont {J.~A.}\ \bibnamefont
  {Blackmore}}, \bibinfo {author} {\bibfnamefont {L.}~\bibnamefont {Caldwell}},
  \bibinfo {author} {\bibfnamefont {P.~D.}\ \bibnamefont {Gregory}}, \bibinfo
  {author} {\bibfnamefont {E.~M.}\ \bibnamefont {Bridge}}, \bibinfo {author}
  {\bibfnamefont {R.}~\bibnamefont {Sawant}}, \bibinfo {author} {\bibfnamefont
  {J.}~\bibnamefont {Aldegunde}}, \bibinfo {author} {\bibfnamefont
  {J.}~\bibnamefont {Mur-Petit}}, \bibinfo {author} {\bibfnamefont
  {D.}~\bibnamefont {Jaksch}}, \bibinfo {author} {\bibfnamefont {J.~M.}\
  \bibnamefont {Hutson}}, \bibinfo {author} {\bibfnamefont {B.~E.}\
  \bibnamefont {Sauer}}, \bibinfo {author} {\bibfnamefont {M.~R.}\ \bibnamefont
  {Tarbutt}}, \ and\ \bibinfo {author} {\bibfnamefont {S.~L.}\ \bibnamefont
  {Cornish}},\ }\bibfield  {title} {\enquote {\bibinfo {title} {Ultracold
  molecules for quantum simulation: rotational coherences in {CaF} and
  {RbCs}},}\ }\href {\doibase 10.1088/2058-9565/aaee35} {\bibfield  {journal}
  {\bibinfo  {journal} {Quantum Sci. Technol.}\ }\textbf {\bibinfo {volume}
  {4}},\ \bibinfo {pages} {014010} (\bibinfo {year} {2018})}\BibitemShut
  {NoStop}%
\bibitem [{\citenamefont {Childs}\ \emph
  {et~al.}(1981{\natexlab{b}})\citenamefont {Childs}, \citenamefont {Goodman},\
  and\ \citenamefont {Goodman}}]{Childs:JMS1981:CaF}%
  \BibitemOpen
  \bibfield  {author} {\bibinfo {author} {\bibfnamefont {W.~J.}\ \bibnamefont
  {Childs}}, \bibinfo {author} {\bibfnamefont {G.~L.}\ \bibnamefont {Goodman}},
  \ and\ \bibinfo {author} {\bibfnamefont {L.~S.}\ \bibnamefont {Goodman}},\
  }\bibfield  {title} {\enquote {\bibinfo {title} {{Precise determination of
  the $\nu$ and $N$ dependence of the spin-rotation and hyperfine interactions
  in the CaF X$^2\Sigma_{1/2}$ ground state}},}\ }\href
  {https://doi.org/10.1016/0022-2852(81)90288-5} {\bibfield  {journal}
  {\bibinfo  {journal} {J. Mol. Spectrosc.}\ }\textbf {\bibinfo {volume}
  {86}},\ \bibinfo {pages} {365--392} (\bibinfo {year}
  {1981}{\natexlab{b}})}\BibitemShut {NoStop}%
\bibitem [{\citenamefont {Tarbutt}\ \emph {et~al.}(2009)\citenamefont
  {Tarbutt}, \citenamefont {Hudson}, \citenamefont {Sauer},\ and\ \citenamefont
  {Hinds}}]{Tarbutt2009}%
  \BibitemOpen
  \bibfield  {author} {\bibinfo {author} {\bibfnamefont {M.~R.}\ \bibnamefont
  {Tarbutt}}, \bibinfo {author} {\bibfnamefont {J.~J.}\ \bibnamefont {Hudson}},
  \bibinfo {author} {\bibfnamefont {B.~E.}\ \bibnamefont {Sauer}}, \ and\
  \bibinfo {author} {\bibfnamefont {E.~A.}\ \bibnamefont {Hinds}},\ }\bibfield
  {title} {\enquote {\bibinfo {title} {{Prospects for measuring the electric
  dipole moment of the electron using electrically trapped polar molecules}},}\
  }\href {\doibase 10.1039/b820625b} {\bibfield  {journal} {\bibinfo  {journal}
  {Faraday Discuss.}\ }\textbf {\bibinfo {volume} {142}},\ \bibinfo {pages}
  {37} (\bibinfo {year} {2009})}\BibitemShut {NoStop}%
\bibitem [{\citenamefont {Caldwell}\ \emph {et~al.}(2019)\citenamefont
  {Caldwell}, \citenamefont {Devlin}, \citenamefont {Williams}, \citenamefont
  {Fitch}, \citenamefont {Hinds}, \citenamefont {Sauer},\ and\ \citenamefont
  {Tarbutt}}]{Caldwell2019}%
  \BibitemOpen
  \bibfield  {author} {\bibinfo {author} {\bibfnamefont {L.}~\bibnamefont
  {Caldwell}}, \bibinfo {author} {\bibfnamefont {J.~A.}\ \bibnamefont
  {Devlin}}, \bibinfo {author} {\bibfnamefont {H.~J.}\ \bibnamefont
  {Williams}}, \bibinfo {author} {\bibfnamefont {N.~J.}\ \bibnamefont {Fitch}},
  \bibinfo {author} {\bibfnamefont {E.~A.}\ \bibnamefont {Hinds}}, \bibinfo
  {author} {\bibfnamefont {B.~E.}\ \bibnamefont {Sauer}}, \ and\ \bibinfo
  {author} {\bibfnamefont {M.~R.}\ \bibnamefont {Tarbutt}},\ }\bibfield
  {title} {\enquote {\bibinfo {title} {{Deep Laser Cooling and Efficient
  Magnetic Compression of Molecules}},}\ }\href {\doibase
  10.1103/PhysRevLett.123.033202} {\bibfield  {journal} {\bibinfo  {journal}
  {Phys. Rev. Lett.}\ }\textbf {\bibinfo {volume} {123}},\ \bibinfo {pages}
  {033202} (\bibinfo {year} {2019})}\BibitemShut {NoStop}%
\bibitem [{\citenamefont {Cheuk}\ \emph {et~al.}(2018)\citenamefont {Cheuk},
  \citenamefont {Anderegg}, \citenamefont {Augenbraun}, \citenamefont {Bao},
  \citenamefont {Burchesky}, \citenamefont {Ketterle},\ and\ \citenamefont
  {Doyle}}]{Cheuk2018}%
  \BibitemOpen
  \bibfield  {author} {\bibinfo {author} {\bibfnamefont {L.~W.}\ \bibnamefont
  {Cheuk}}, \bibinfo {author} {\bibfnamefont {L.}~\bibnamefont {Anderegg}},
  \bibinfo {author} {\bibfnamefont {B.~L.}\ \bibnamefont {Augenbraun}},
  \bibinfo {author} {\bibfnamefont {Y.}~\bibnamefont {Bao}}, \bibinfo {author}
  {\bibfnamefont {S.}~\bibnamefont {Burchesky}}, \bibinfo {author}
  {\bibfnamefont {W.}~\bibnamefont {Ketterle}}, \ and\ \bibinfo {author}
  {\bibfnamefont {J.~M.}\ \bibnamefont {Doyle}},\ }\bibfield  {title} {\enquote
  {\bibinfo {title} {{$\Lambda$ -Enhanced Imaging of Molecules in an Optical
  Trap}},}\ }\href {\doibase 10.1103/PhysRevLett.121.083201} {\bibfield
  {journal} {\bibinfo  {journal} {Phys. Rev. Lett.}\ }\textbf {\bibinfo
  {volume} {121}},\ \bibinfo {pages} {083201} (\bibinfo {year}
  {2018})}\BibitemShut {NoStop}%
\bibitem [{\citenamefont {Bergeman}\ \emph {et~al.}(1987)\citenamefont
  {Bergeman}, \citenamefont {Erez},\ and\ \citenamefont
  {Metcalf}}]{Bergeman1987}%
  \BibitemOpen
  \bibfield  {author} {\bibinfo {author} {\bibfnamefont {T.}~\bibnamefont
  {Bergeman}}, \bibinfo {author} {\bibfnamefont {G.}~\bibnamefont {Erez}}, \
  and\ \bibinfo {author} {\bibfnamefont {H.~J.}\ \bibnamefont {Metcalf}},\
  }\bibfield  {title} {\enquote {\bibinfo {title} {{Magnetostatic trapping
  fields for neutral atoms}},}\ }\href {\doibase 10.1103/PhysRevA.35.1535}
  {\bibfield  {journal} {\bibinfo  {journal} {Phys. Rev. A}\ }\textbf {\bibinfo
  {volume} {35}},\ \bibinfo {pages} {1535--1546} (\bibinfo {year}
  {1987})}\BibitemShut {NoStop}%
\bibitem [{\citenamefont {Kajita}(2008)}]{Kajita2008}%
  \BibitemOpen
  \bibfield  {author} {\bibinfo {author} {\bibfnamefont {M.}~\bibnamefont
  {Kajita}},\ }\bibfield  {title} {\enquote {\bibinfo {title} {{Prospects of
  detecting $m_e$/$m_p$ variance using vibrational transition frequencies of
  $^2\Sigma$-state molecules}},}\ }\href {\doibase 10.1103/PhysRevA.77.012511}
  {\bibfield  {journal} {\bibinfo  {journal} {Phys. Rev. A}\ }\textbf {\bibinfo
  {volume} {77}},\ \bibinfo {pages} {012511} (\bibinfo {year}
  {2008})}\BibitemShut {NoStop}%
\bibitem [{zen()}]{zenodo}%
  \BibitemOpen
  \href@noop {} {}\bibinfo {note}
  {\url{https://doi.org/10.5281/zenodo.3382223}}\BibitemShut {NoStop}%
\end{thebibliography}%

\end{document}


\title{Long rotational coherence times of molecules in a magnetic trap: \\Supplemental Material}

\author{L. Caldwell}
\affiliation{Centre for Cold Matter, Blackett Laboratory, Imperial College
London, Prince Consort Road, London SW7 2AZ UK}

\author{H. J. Williams}
\affiliation{Centre for Cold Matter, Blackett Laboratory, Imperial College
London, Prince Consort Road, London SW7 2AZ UK}

\author{N. J. Fitch}
\affiliation{Centre for Cold Matter, Blackett Laboratory, Imperial College
London, Prince Consort Road, London SW7 2AZ UK}

\author{J. Aldegunde}
\affiliation{Departamento de Quimica Fisica, Universidad de Salamanca, E-37008,
Salamanca, Spain}
\author{Jeremy M. Hutson}
\affiliation{Joint Quantum Centre (JQC) Durham-Newcastle, Department of
Chemistry, Durham University, South Road, Durham DH1 3LE, UK}

\author{B. E. Sauer}
\affiliation{Centre for Cold Matter, Blackett Laboratory, Imperial College
London, Prince Consort Road, London SW7 2AZ UK}

\author{M. R. Tarbutt}
\affiliation{Centre for Cold Matter, Blackett Laboratory, Imperial College
London, Prince Consort Road, London SW7 2AZ UK}

\date{\today}


\maketitle
\section{Electronic structure calculations of $g_\textrm{r}$ and $g_l$}

We calculate $g_l$, the anisotropy of the electronic $g$ tensor, using density
functional theory (DFT) with the B3LYP functional \cite{Stephens:1994}. This
functional was chosen because, as discussed in reference
\cite{aldegunde:2Sigma:pra2019}, it is better than other functionals at
reproducing experimental values of $g_l$ and $\gamma$; the latter characterizes
the electron spin-rotation interaction and is connected to $g_l$ via Curl's
approximation \cite{curl:mp1965}. The calculations include relativistic
corrections via the zero-order regular approximation (ZORA)
\cite{vanLenthe:1993, vanLenthe:1994,vanLenthe:1999} and are performed with the
ORCA code \cite{Neese:2012}. Experimental values of $g_l$ are available for
numerous $^2\Sigma$ molecules, including most of those discussed in Table~\ref{tab:glgr} \cite{Anderson:JCP1994:MgF,Childs:JMS1981:CaF,Childs:JMS1981:SrF,Knight:JCP1972:BeH,Bernath:TAJ1985:MgH,Berg1974:CaH,:SrH,Brom:JCP1976:BeOH,Brom:JCP:1973:MgOH,Ziurys:TAJ1992:CaOH,Anderson:CPL1992:SrOH}. In
general, our calculations agree with experimental values within 5\%, although
the difference may increase up to 20\% when comparing with measurements
performed in noble-gas matrices or when the value of $g_l$ is not directly
measured but is extracted from $\gamma$ using Curl's approximation, which is
known to introduce an error around 15\% \cite{Weltner:book}. Such larger
discrepancies occur, for example, for MgOH (measured in a matrix \cite{Brom:JCP:1973:MgOH}) and CaH
($g_l$ obtained from $\gamma$ \cite{Berg1974:CaH}).

We are unable to use the same methodology for the rotational $g$ factor,
$g_\textrm{r}$, because its calculation is not supported by ORCA. As
correlation effects are relatively small for $g_\textrm{r}$
\cite{Mohn:AQC2005}, we calculate it using Hartree-Fock (HF) calculations
performed with the DALTON code \cite{Dalton:current}. Unfortunately, very few
experimental values of $g_\textrm{r}$ are available for open-shell molecules
and, in particular, for those included in Table~\ref{tab:glgr}. The only exception to this
is CaF, for which the calculated value is in good agreement with the
experimental value obtained in the present work, $g_\textrm{r} =
-5.0(2)\times 10^{-5}$.

We initially attempted to perform calculations on Yb as well as the lighter metals in Table~\ref{tab:glgr},
but ultimately decided they were not reliable enough for publication. To provide consistency across all the elements, we chose to use the x2c-TZVPP basis sets of Pollak and Weigend \cite{Pollak:2017}
for both DFT and HF calculations. These basis sets are of triple-$\zeta$
quality, and are designed for use in relativistic calculations. They are
available for all atoms in the periodic table up to Rn.

The DFT and HF methods we use here are reliable only for molecules that are
reasonably well described by a single electron configuration. In order to
verify this, we performed CCSD(T) calculations and used the T1 diagnostic
\cite{Lee:T1} to assess the multiconfigurational character of the ground
electronic state. A value $\textrm{T1}>0.02$ is generally considered to
indicate that multireference calculations are needed. For the molecules in
Table~\ref{tab:glgr}, T1 was below this threshold, indicating that the ground state is
reasonably well described by a single configuration.
\newbibstartnumber{33}
\bibliography{Sup_Mat}